\documentclass[iop]{emulateapj}
\usepackage{graphicx}
\usepackage{mdwlist}
\usepackage{url}
\usepackage[bookmarks=false]{hyperref}

\newcommand{\tnt}{\tablenotetext}
\newcommand{\tnm}{\tablenotemark}
\newcommand{\HBootes}{HBo\"{o}tes}

\begin{document}

\shorttitle{SEDs of Six DSFGs}
\shortauthors{Ma et al.}

\title{Spitzer Imaging of Strongly-Lensed Herschel-Selected Dusty Star Forming Galaxies}
\author{Brian Ma\altaffilmark{1},
Asantha Cooray\altaffilmark{1,2},
J. A. Calanog\altaffilmark{1},
H. Nayyeri\altaffilmark{1},
N. Timmons\altaffilmark{1},
C. Casey\altaffilmark{1},
M. Baes\altaffilmark{3},
S. Chapman\altaffilmark{4},
H. Dannerbauer\altaffilmark{5},
E. L. Da Cunha\altaffilmark{6},
G. De Zotti\altaffilmark{7,8},
L. Dunne\altaffilmark{9,10},
D. Farrah\altaffilmark{11},
Hai Fu\altaffilmark{12},
J. Gonzalez-Nuevo\altaffilmark{13},
G. Magdis\altaffilmark{14},
M.~J.~Micha\l{}owski\altaffilmark{9},
I. Oteo\altaffilmark{9,15},
D. A. Riechers\altaffilmark{16},
D. Scott\altaffilmark{17},
M. W. L. Smith\altaffilmark{18},
L. Wang\altaffilmark{19,20},
J.~Wardlow\altaffilmark{21},
M.~Vaccari\altaffilmark{22},
S. Viaene\altaffilmark{3},
J. D. Vieira\altaffilmark{23},
and
M. Vaccari\altaffilmark{24}
}
\altaffiltext{1}{Department of Physics and Astronomy, University of California, Irvine, CA 92697, USA}
\altaffiltext{2}{California Institute of Technology, 1200 E. California Blvd., Pasadena, CA 91125, USA}
\altaffiltext{3}{Sterrenkundig Observatorium, Universiteit Gent, Krijgslaan 281 S9, B-9000 Gent, Belgium}
\altaffiltext{4}{Department of Physics and Atmospheric Science, Dalhousie University, Halifax, Nova Scotia, B3H 4R2, Canada}
\altaffiltext{5}{Laboratoire AIM-Paris-Saclay, CEA/DSM/Irfu-CNRS-Universit\'{e} Paris Diderot, CE-Saclay, pt courrier 131, F-91191 Gif-sur-Yvette, France}
%\altaffiltext{5}{Universitt\"at Wien, Institt f\"ur Astrophysik, T\"urkenschanzstra{\ss}e 17, 7, 1180 Wien, Aust}
\altaffiltext{6}{Center for Astrophysics and Supercomputing, Swinburne University of Technology, Hawthorn VIC 3122, Australia}
\altaffiltext{7}{INAF-Osservatorio Astronomico di Padova, Vicolo Osservatorio 5, I-35122 Padova, Italy}
\altaffiltext{8}{SISSA, Via Bonomea 265, I-34136 Trieste, Italy}
\altaffiltext{9}{Institute for Astronomy, University of Edinburgh, Royal Observatory, Blackford Hill, Edinburgh, EH9 3HJ, United Kingdom}
\altaffiltext{10}{Department of Physics and Astronomy, University of Canterbury, Private Bag 4800, Christchurch, 8140, New Zealand}
\altaffiltext{11}{Department of Physics, Virginia Tech, Blacksburg, VA 24061, USA}
\altaffiltext{12}{Department of Physics and Astronomy, University of Iowa, Van Allen Hall, Iowa City, IA 52242, USA}
\altaffiltext{13}{Departamento de Fisica, Universidad de Oviedo C/ Calvo Sotelo, s/n, 33007 Oviedo, Spain}
\altaffiltext{14}{Department of Astrophysics, Denys Wilkinson Building, University of Oxford, Keble Road, Oxford OX1 3RH, United Kingdom}
\altaffiltext{15}{European Southern Observatory, Karl-Schwarzschild-Str. 2, 85748 Garching, Germany}
\altaffiltext{16}{Department of Astronomy, Cornell University, 220 Space Sciences Building, Ithaca, NY 14853, USA}
\altaffiltext{17}{Department of Physics and Astronomy, University of British Columbia, 6224 Agricultural Road, Vancouver, BC V6T 1Z1, Canada}
\altaffiltext{18}{Department of Physics and Astronomy, Cardiff University, Cardiff, CF24 3AA, United Kingdom}
\altaffiltext{19}{Department of Physics, Institute for Computational Cosmology, Durham University, Durham, County Durham DH1, United Kingdom}
\altaffiltext{20}{SRON Netherlands Institute for Space Research, Landleven 12, 9747 AD, Groningen, The Netherlands}
\altaffiltext{21}{Dark Cosmology Centre, Niels Bohr Institute, University of Copenhagen, Denmark}
\altaffiltext{22}{Physics Department, University of the Western Cape Private Bag X17, 7535, Bellville, Cape Town, South Africa}
\altaffiltext{23}{Department of Physics and Astronomy, University of Illinois, 1002 W. Green St., Urbana, IL 61801, USA}
\altaffiltext{24}{Astrophysics Group Physics Department, University of the Western Cape, Private Bag X17, 7535, Bellville, Cape Town, South Africa}

%%%%% Abstract %%%%%
\begin{abstract}

We present the rest-frame optical spectral energy distribution and stellar masses of six {\it Herschel}-selected gravitationally lensed dusty, star-forming galaxies (DSFGs) at $1<z<3$. These galaxies were first identified with {\it Herschel}/SPIRE imaging data from the {\it Herschel} Astrophysical Terahertz Large Area Survey (H-ATLAS) and the {\it Herschel} Multi-tiered Extragalactic Survey (HerMES). The targets were observed with {\it Spitzer}/IRAC at 3.6 and 4.5$\mu$m. Due to the spatial resolution of the IRAC observations at the level of 2$''$, the lensing features of a background DSFG in the near-infrared are blended with the flux from the foreground lensing galaxy in the IRAC imaging data. We make use of higher resolution {\it Hubble}/WFC3 or Keck/NIRC2 Adaptive Optics imaging data to fit light profiles of the foreground lensing galaxy (or galaxies) as a way to model the foreground components, in order to successfully disentangle the foreground lens and background source flux densities in the IRAC images. The flux density measurements at 3.6 and 4.5$\mu$m, once combined with {\it Hubble}/WFC3 and Keck/NIRC2 data, provide important constraints on the rest-frame optical spectral energy distribution of the {\it Herschel}-selected lensed DSFGs. We model the combined UV- to millimeter-wavelength SEDs to establish the stellar mass, dust mass, star-formation rate, visual extinction, and other parameters for each of these {\it Herschel}-selected DSFGs. These systems have inferred stellar masses in the range $8\times10^{10}$ to $4\times10^{11}$ M$_{\odot}$ and star-formation rates of around 100 M$_{\odot}$ yr$^{-1}$. This puts these lensed sub-millimeter systems well above the SFR-\textit{M$^*$} relation observed for normal star-forming galaxies at similar redshifts. The high values of SFR inferred for these systems are consistent with a major merger-driven scenario for star formation. 
\end{abstract}

%%%%% Introduction %%%%%
\section{Introduction}

	Dusty star-forming galaxies (DSFGs; For a recent review, see \citealt{Casey2014}) are now believed to significant contributors to cosmic star-formation in the early Universe (e.g. \citealt{LeFloch2005,PerezGonzalez2005,Takeuchi2005,Marchesini2014}). The extreme star-bursting examples of DSFGs appear as bright sub-millimeter galaxies (SMGs, see \citealt{Smail1997,Hughes1998,Barger1998,Coppin2008,Austermann2009}) and are best studied at far-IR and sub-millimeter wavelengths due to the high dust extinction at rest-frame optical wavelengths.  Such galaxies have star-formation rates in excess of 100 M$_{\sun}$ yr$^{-1}$ and have emissions peaking in the far-infrared, with luminosities $L_{\rm FIR} > 10^{12}$ L$_{\sun}$. The strong clustering of these starburst galaxies \citep{Blain2004,Farrah2006,Cooray2010,Hickox2012} is such that they may evolve into dark matter halos that host some of the most luminous and massive elliptical galaxies today. Thus their rapid formation may have moved these galaxies to the red sequence that already exists for galaxies at $z > 1$ \citep{Faber2007,Barro2013,Ilbert2013}. It is generally believed that a large fraction of the high-redshift DSFGs are trigged by galaxy mergers, similar to local ULIRGS \citep{Sanders1996}. There is also observational evidence for lack of merger features in the optical and infrared images in $>40$\% of the DSFGs at $z \sim 1$ \citep{Kartaltepe2011}.

	Despite significant progress in understanding the formation and evolution of DSFGs at $z > 1$, it still remains difficult to measure basic properties, such as the total amount of gas, stars, and dust in these objects, as well as the spatial distribution of these quantities within the galaxies. The primary obstacles for such detailed observations come from the poor spatial resolution of the far-infrared/sub-millimeter observations with which the DSFGs are identified, and their faintness at near-infrared wavelengths. The former makes it challenging to identify their counterparts at other wavelengths, while the latter makes follow-up observations time-consuming. Nevertheless, the rest-frame near-IR flux densities of galaxies are crucial to establish the total stellar mass content and to study the spatial distribution of stellar populations in such objects. While these studies may be challenging for typical dusty starbursts, gravitationally lensed DSFGs provide a mechanism to partially overcome the limitation associated with the faintness of the near-IR counterparts, specifically through the flux enhancement associated with lensing magnification. Lensed DSFGs also provide enhanced spatial resolution through lensing magnification, and this enhancement has provided information down to 200-300pc scales within the interstellar medium of some of the lensed SMGs, such as SMMJ2135-0102 of \citet{Swinbank2010} and SDP.81 of \citet{Negrello2010} that has been studied with an extended observation using long baselines with ALMA \cite{Dye2015,Alma2015}.

	Despite progress in high-resolution imaging in the mm-wavelengths with interferometers and large optical and infrared telescopes in the rest-frame wavelengths, our views of distant galaxies at wavelengths above 2.2 $\mu$m and above in the mid-infrared region are still limited to imaging data that are diffraction limited. For galaxies at $z\sim 1$ to 3, such data are still crucial since they provide necessary information to break certain degeneracies in models of the spectral energy distribution (SED). A key ingredient from SED analysis is the stellar mass of the galaxies, and with data out to 2.2 $\mu$m only, it is generally hard to accurately estimate the stellar mass. One issue is that due to high extinction, many of the DSFGs are faint in the rest-frame UV observable at wavelengths below 1 $\mu$m. Despite limitations, {\it Spitzer}/IRAC data remain key to estimating the stellar masses of DSFGs, as past studies have demonstrated \citep{Hopwood10,Michalowski2012,Michalowski2014}.

	In this paper we discuss the rest-frame optical to sub-millimeter spectral energy distributions of six lensed DSFGs. Our key measurements in the optical to infrared are with {\it Spitzer}/IRAC at $z=1$ to 3, which are crucial to improve the accuracies of stellar mass measurements using models of the spectral energy distributions. Due to the spatial resolution of IRAC imaging at the level of 2$''$ per pixel, the lensed DSFGs are blended with their respective foreground lensing galaxies in our IRAC images. The paper is organized as follows. In the next Section we describe the target selection and the observations we have acquired. In Section~\ref{sec:deblend} we outline the procedure that was used to de-blend the background DSFGs from the foreground lenses in the IRAC images. In Section~\ref{sec:sed} we discuss the SEDs of these galaxies and present model estimates of stellar mass, star-formation rate, dust mass, and other quantities. We conclude with a discussion of our results in Section~\ref{sec:results}.

	Throughout this paper we assume a Chabrier (2003) initial mass function (IMF) with a cutoff below 0.1 and above 100 M$_{\odot}$. We assume a ${\Lambda}$CDM cosmology with $H_{0} = 70~\rm{km~s^{-1}~Mpc}^{-1}$, $\Omega_{m} = 0.3$, and $\Omega_{\Lambda} = 0.7$.

%%%%%Target Selection and Observations%%%%%
\section{Target Selection and Observations}

	 The sample of lensed DSFGs studied in this paper was originally selected from two key surveys, H-ATLAS \citep{Eales2010} and HerMES \citep{Oliver2012} which were completed with the {\it Herschel} Space Observatory \citep{Pilbratt2010}. The lensing selection at sub-millimeter wavelengths simply involves a flux density cut at 500$\mu$m and an accounting for bright nearby galaxies or galaxies that harbor radio-loud AGN \citep{Negrello2010,Wardlow2013}. The HerMES lensing sample (see Riechers et al., in prep. for a study on CO redshifts) and its selection is discussed in \citet{Wardlow2013}, while the H-ATLAS sample is discussed in \citet{Negrello2010}, \citet{Bussmann2013}, and \citet{Calanog2014}.
	 
	With our available {\it Spitzer}/IRAC data of several DSFGs, we select those that were classified as ``Grade A'' lenses following the designation by \citet{Calanog2014} as our sources for study. These sources have visually obvious lensing features as observed in near-IR data, either with {\it HST}/WFC3 or Keck/NIRC2 Adaptive Optics imaging in the $K_s$ 2.2$\mu$m band. Lensing morphologies include rings, arcs, and counter-images detected at high-significance. Some sources (HFLS08 and HLock12 in this study) are also classified as Grade A if a possible counter-image can be properly modeled after subtracting the foreground lens in the higher resolution data. As an extra check, DSFGs are classified as Grade A if their near-IR morphologies resemble the configurations found in high-resolution sub-millimeter data \citep{Bussmann2013}. Because this designation is based off detections in the near-infrared, there is a selection bias towards higher stellar masses. Our sample of six DSFGs includes HATLASJ114638.0-001132 (G12v2.30) from \citet{Fu2012} and HATLASJ142935.3-002836 (G15v2.19) from \citet{Messias2014} and \citet{Timmons2015}. 

%\input{./Tables/hires}
%Table summarizing high-resolution data
\begin{deluxetable*}{l c c c c}
        \tablecolumns{4}
        \tablewidth{\linewidth}
        \tablecaption{Summary of High Resolution Data}
        \tablehead{
                \colhead{IAU Name} &
                \colhead{Short Name} &
                \colhead{$z_{\rm source}$} &
                \colhead{Ref.} &
                \colhead{Exp. Time}
                \\
                \colhead{} &
                \colhead{} &
                \colhead{} &
                \colhead{} &
                \colhead{Filter = $t_{\rm int}\tnm{$^1$} \times N_{\rm frames}\tnm{$^2$}$}
        }
        \startdata
        
        1HerMES S250 J142825.7+345547  & \HBootes02 & 2.804 & R14 & $J_{\rm F110W}=62 \times 4, H=120 \times 28, K_{\rm s}=80 \times 27$ \\
        1HerMES S250 J171544.9+601239 & HFLS08 & 2.264 & R14 & $J_{\rm F110W}=62 \times 4$ \\
        HATLASJ085358.9+015537 & G09v1.40 & 2.091 & L14 & $K_{\rm s}=80 \times 45$ \\
        HATLASJ114638.0-001132 & G12v2.30 & 3.259 & H14 & $K_{\rm s}=80 \times 42$ \\
        HATLASJ142935.3-002836 & G15v2.19 & 1.026 & M14 & $J_{\rm F105W}=88 \times 4, J_{\rm F160W}=62 \times 4, H=120 \times 10, K_{\rm s}=80 \times 15$ \\
        1HerMES S250 J110016.3+571736 & HLock12 & 1.651 & R14 & $J_{\rm F110W}=62 \times 4$

        \enddata
        \label{tab:hires}
        \tablecomments{Redshift key: R14 = Riechers et al. (in prep); L14 = R. E. Lupu et al. (in prep); H14 = A. I. Harris et al. (in prep); and M14 = \citet{Messias2014}. Filters are $J_{F105W}=$ {\it HST} F105W, $J_{\rm F110W}=$ {\it HST} F110W, $J_{\rm F160W}=$ {\it HST} F160W, $H=$ Keck {\it H}-band, and $K_{\rm s}=$ Keck $K_{s}$-band.}
        \tnt{$^1$}{$t_{\rm int}$ is the exposure time per frame in seconds}
        \tnt{$^2$}{$N_{\rm frames}$ is the number of independent frames}
\end{deluxetable*}

\subsection{HST: WFC3}
	\textit{Herschel}-lensing candidates in the H-ATLAS and HerMES fields were observed as part of the {\it HST} WFC3 Cycle 19 program (P.I. M. Negrello). Images were taken with the F110W filter ($\lambda_{\rm c}=1.15 {\mu}$m), with integration times typically $\sim$ 4 minutes for each target, with a depth of 25.4 AB magnitude. We also include additional imaging observed through the F105W ($\lambda_{\rm c}=1.06 {\mu}$m) and F160W ($\lambda_{\rm c}=1.54 {\mu}$m) filters for HATLASJ142935.3-002836 that were carried out as part of a grism observation of that galaxy, with grism data reported in \citet{Timmons2015}. {\it HST}/WFC3 data reduction procedures are described in \citet{Calanog2014}.

\subsection{Keck: NIRC2 LGS-AO}
	Keck imaging data in the $K_{\rm s}$-band was obtained using the NIRC2 LGS-AO system during the years 2011 to 2013. Due to LGS-AO observations, these imaging data have spatial resolutions comparable to or better than {\it HST}/WFC3 imaging data, reaching AO-corrected resolutions of 0.1$''$ in the best conditions.
	Exposure times were usually 45 minutes to obtain a $5\sigma$ point source depth of 25.7 AB magnitude using a $0.1''$ aperture radius \citep{Calanog2014}. Data reduction was carried out with an internal pipeline code that is described in \citet{Fu2012,Fu2013}. Exposure times for high-resolution data are presented in Table \ref{tab:hires}.
	
\begin{figure}
	\begin{minipage}{\linewidth}
		\hspace{-1.5cm}
		\includegraphics[scale=0.6]{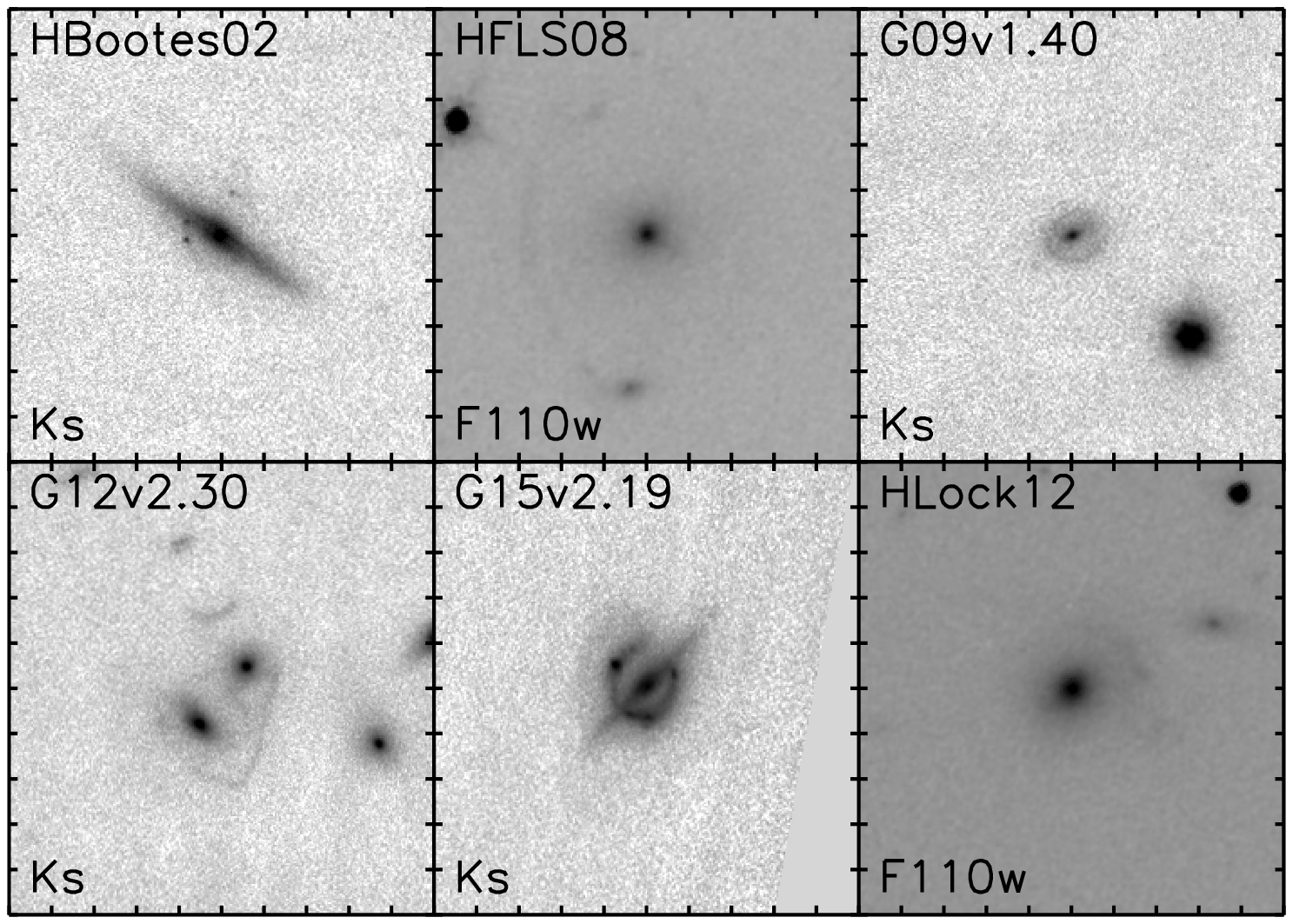}
		\vspace{-0.75cm}
		\caption{$10''\times10''$ high-resolution data of selected DSFGs through various filters. Each image is oriented north upwards and east to the left, with tickmarks representing 1$''$.}
		\label{fig:hires}
	\end{minipage}
\end{figure}

\subsection{Spitzer IRAC}

	The six lensed DSFGs was observed with {\it Spitzer}/IRAC in Cycle 8 through IRAC channels 1 ($\lambda_{\rm c} = 3.6\mu$m) and 2 ($\lambda_{\rm c} = 4.5\mu$m) as part of program 80156 (P.I. A. Cooray). Most targets were observed through 72 individual tiles using a 12-point box dithering pattern with 30 seconds of exposure each, for a total exposure time of 36 minutes. The exception is 1HerMES S250 J142825.7+345547, which was observed through 48 tiles for 24 minutes in total. We used the MOsaicker and Point source EXtractor (MOPEX, v18.5.0) program on basic calibrated data to carry out image reduction and mosaicking. Background estimations were carried out using SExtractor \citep{Bertin1996}. Final images were re-sampled to a pixel size of $0.6''\times0.6''$, with resolutions of 1.65$''$ in FWHM for IRAC 3.6$\mu$m and 1.80$''$ in FWHM for IRAC 4.5$\mu$m.

%%%%% Deblending %%%%%
\section{Deblending Spitzer IRAC Images}
\label{sec:deblend}

	Due to 2$''$ or worse spatial resolution provided by IRAC imaging, the observed DSFGs remain mostly unresolved in comparison to the high-resolution data (FWHM value of $\sim1.8''$ for IRAC imaging compared to $\sim0.2''$ for our {\it HST} and Keck data). This makes modeling the foreground and background sources difficult, due to significant degeneracies between the flux distribution of the two components. Two previous studies on lensed DSFGs observed with {\it Spitzer}/IRAC de-blended the background source flux from that of the foreground lens in different ways. \citet{Hopwood10} used \texttt{galfit} \citep{Peng2010} directly on IRAC images to model the foreground lens. However, attempts to implement this procedure on our images were largely unsuccessful; the models were highly degenerate, primarily because the effective radius of the foreground components  ($\sim$ 1--2 pixels or $\sim$ 2--4'') often overlapped with the background source. The second approach---and the method that this study uses---is to convolve \texttt{galfit} models of high-resolution near-IR data with the IRAC beam, rescale the model into the same pixel scale as the IRAC image, and subtract it from the IRAC data \citep{Bothwell2013}. We make use of the higher resolution HST and/or Keck/NIRC2 Adaptive Optics imaging to de-blend the lens flux density from that of the background DSFG in our IRAC data. The fundamental assumption of this method is that these models retain the same morphologies across near-IR wavelengths from 1.1$\mu$m to 4.5$\mu$m. For galaxies at $z \sim 2$ to 3 this assumption could have some issues since the 1.1$\mu$m images are below the 4000\AA\ break while 3.6$\mu$m image is sampling redward of the break in the rest-frame optical. This is less of an issue once the $K_{\rm s}$-band images are combined with IRAC imaging, since in these cases all of the images sample the galaxy redward of the 4000\AA\ break. We have two targets without ground-based $K_{\rm s}$-band imaging using the Keck/LGS-AO system, due to the lack of suitable AO guide stars within 45--60$''$ of their location. In those two cases, unfortunately we are forced to use the HST/WFC3 1.1$\mu$m image as that provides the highest resolution imaging of the system.

	As the Keck and HST/WFC3 data of all our targets were introduced and studied in \citet{Calanog2014} we used high-resolution models developed 
there for the de-blending process. Best-fit foreground models were obtained from \texttt{galfit}, while models for the background sources were found using \texttt{gravlens} \citep{Keeton01}. HATLASJ114638.0-001132 has been studied extensively in previous works, and we used models provided by \citet{Fu2012}.  An empirical PSF from each IRAC image was obtained by stacking $\sim$ 10 bright and unsaturated sources in the same image, and then taking the average. Since the IRAC PSF is triangular, we make sure the PSF and the IRAC images are oriented the same before convolution with the best-fit models. We normalized the PSF by the ratio of the peak high- to low-resolution fluxes. The beam-convolved near-IR models were then rescaled to the same pixel scale as the IRAC data and subtracted from the IRAC images. 
	
	In order to obtain optimal fits and subtractions, we scaled the foreground and background models simultaneously and corrected for positional errors (i.e., we fit for separate scale factors for both models, as well as positional shifts in the horizontal and vertical directions) using the iterative IDL routine \texttt{amoeba\_sa}, which utilizes the downhill simplex method in multi-dimensions to minimize multi-dimensional functions. During the fitting, we allow for a maximum positional shift of 1 pixel to prevent accidental fitting of tertiary sources, an issue that is especially prevalent in crowded fields, as found for HFLS08 and G12v2.30. Uncertainties in the scale factors and position shifts were calculated using the Markov Chain Monte Carlo technique (MCMC). The convolved models and residuals are shown in the Appendix. 
	
	We determine the goodness of the fits based on the minimized, reduced $\chi^{2}_{\nu}$ statistic, as given by:
	
	\begin{equation}
		\chi^{2}_{\nu} = \frac{1}{N_{DOF}}\sum\limits_{x=1}^{nx}\sum\limits_{y=1}^{ny}\frac{(f_{\rm data}(x,y)-f_{\rm model}(x,y))^{2}}{\sigma(x,y)^{2}},
	\end{equation}
\noindent	
where $f_{\rm model}$ is the sum of the convolved foreground and background models, as well as any secondary components (used on a case-by-case basis, see Section \ref{sec:notes}), and $N_{\rm DOF}$ is the number of degrees of freedom (number of pixels $-$ number of free parameters).
	
\subsection{Notes on Individual Sources}
\label{sec:notes}
	We now briefly discuss each lensed DSFG as well as their individual de-blending solutions. These results are shown in the Appendix.
	\newline
	\newline
	\textbf{1HerMES S250 J142825.7+345547 (\HBootes02):} The lensing morphology in high-resolution imaging shows an Einstein ring lensed by an edge-on foreground galaxy. A well-isolated source in IRAC imaging, de-blending the foreground and background lenses was straightforward. Light profile fits to the lensing galaxy are the most robust and reliable of the sample in this study, with $\chi^2_\nu$ values on the order of unity. The fitting at 4.5$\mu$m was significantly worse than in 3.6$\mu$m, most likely due to poor imaging and consequentially a suboptimal convolution with the PSF. 
	\newline
	\newline
	\textbf{1HerMES S250 J171544.9+601239 (HFLS08):} High-resolution near-IR imaging shows an arc about $3''$ east of the foreground galaxy. Moreover, this target is located in a crowded region, with two other sources a few arcseconds away to the east, whose light could potentially pollute the contributions from the foreground and background galaxies. Attempts to fit for these secondary sources only introduced additional uncertainties. Final convolved models are highly degenerate and the photometry and subsequent analyses of this source should be treated with caution. 
	\newline
	\newline
	\textbf{HATLASJ085358.9+015537 (G09v1.40):} The lensing feature is an Einstein ring centered almost exactly on the foreground lens. The de-blending process for this source is unique, because a separate fit was made for a source located about 2$''$ to the southwest. Fitting for this object was deemed necessary because imperfections in the PSF introduced degeneracies and an increased systematic uncertainty to the light profile modeling. 
	\newline
	\newline
	\textbf{HATLASJ114638.0-001132 (G12v2.30):} High-resolution data has shown the lensing morphology to be highly complex and lensed by four foreground sources in the near-IR---for a detailed analysis, see \citet{Fu2012}. Similar to HFLS08, this field is fairly crowded, with multiple sources within 4$''$ to the southeast and southwest. Attempts to fit for the nearby sources created very degenerate results, so only the foreground and background components were used in the de-blending process. It should be noted that light from neighboring sources may skew photometry of the background DSFG.
	\newline
	\newline
	\textbf{HATLASJ142935.3-002836 (G15v2.19):} The lensing morphology is an incomplete Einstein ring around a star-forming, edge-on spiral galaxy and has been studied extensively in \citet{Messias2014} and \citet{Timmons2015}.  The source is sufficiently isolated so that no light pollution from nearby sources can affect the fitting. Despite this, the final results were not very robust. This raises the possibility that the lensing morphologies observed at 3.6${\mu}$m and 4.5${\mu}$m vary significantly from those observed in the $\sim$ 1--2 $\mu$m range.
	\newline
	\newline
	\textbf{1HerMES S250 J110016.3+571736 (HLock12):} High-resolution near-IR imaging shows that the lensing morphology is an arc approximately $1''$ northwest of the foreground source. Subtraction of the foreground galaxy in high-resolution imaging has also revealed a counter-image less than $1''$ to the southeast of the foreground lens \citep{Calanog2014}. The system lies in a moderately crowded field, so separate fits were made for the sources 4$''$ to the southwest, 4$''$ to the northwest, and 1$''$ to the northeast. Despite the number of sources to fit for, the final model fit and de-blending was found to be robust.

	Throughout the rest of this paper, we shall refer to each galaxy by their respective short names for brevity.
	
%%%%% Photometry %%%%%
\subsection{Photometry}

	We used our own IDL codes to perform aperture photometry on the IRAC background models. Data were converted from [MJy sr$^{-1}$] to [${\mu}$Jy pix$^{-1}$] prior to photometric measurements. We calculated the flux density within a circular aperture of a specified radius that would enclose the DSFG emissions---$2''$ for all sources, except for HFLS08, where we used $4''$, and for G12v2.30 and HLock12, where we used a $3''$ aperture. To find the flux errors, we calculated the flux densities within apertures of the same radii placed at empty areas of the residual image. Finally, we divide the observed flux densities for each source by their respective magnification factors to find their intrinsic flux densities. We assume constant magnification factors in the near-IR wavelengths (i.e., no differential magnification from 1.1$\mu$m imaging data with {\it HST}/WFC3 to {\it Spitzer}/IRAC at 4.5$\mu$m, see \citealt{Calanog2014}). These factors, as well as the far-IR magnification factors, are shown in Table \ref{tab:magfactors}. The measured flux densities and errors can be found with values from other wavelengths in Table~\ref{tab:fluxes}.

%\input{./Tables/magfactors}
%Table of magnification factors used in finding intrinsic fluxes
\begin{deluxetable}{l c c c c c}
        \tablecolumns{3}
        \tablewidth{\linewidth}
        \tablecaption{Magnification Factors of SMGs}
        \tablehead{
                \colhead{} &
                \colhead{${\mu}_{\rm{NIR}}$} & 
                \colhead{Ref.} &
                \colhead{${\mu}_{\rm{FIR}}$} &
                \colhead{Ref.}
        }
        \startdata
        \vspace{0.1cm}
        % HBootes02
        \HBootes02 & $5.3^{+1.4}_{-0.4}$ & C14 & $10.3^{+1.7}_{-1.7}$ & B13
        \\
        \vspace{0.1cm}
        % HFLS08
        HFLS08 & $7.7^{+1.6}_{-0.7}$ & C14 & $8.2$\tnm{$^{1}$} & ---
        \\
        \vspace{0.1cm}
        % G09v1.40
        G09v1.40 & $11.4^{+0.9}_{-1}$ & C14 & $15.3^{+3.5}_{-3.5}$ & B13
        \\
        \vspace{0.1cm}
        % G12v2.30
        G12v2.30 & $16.7^{+0.8}_{-0.8}$ & F12 & $7.6^{+1.5}_{-1.5}$ & F12
        \\
        \vspace{0.1cm}
        % G15v2.19
        G15v2.19 & $9.6^{+1}_{-0.3}$ & C14 & $9.7^{+0.7}_{-0.7}$ & M14
        \\
        \vspace{0.1cm}
        % HLock12
        HLock12 & $4.0^{+0.4}_{-0.4}$ & C14 & $5.5\tnm{$^{1}$}$ & ---
                
        \enddata
        
        \label{tab:magfactors}
        \tablecomments{Magnification factors used to convert from observed to intrinsic flux densities. We use $\mu_{\rm NIR}$ factors to convert 1.1$\mu$m to 4.5$\mu$m flux densities and $\mu_{\rm FIR}$ factors to convert millimeter flux densities. The reference key is: C14 = \citet{Calanog2014}, F12 = \citet{Fu2012}, B13 = \citet{Bussmann2013}, and M14 = \citet{Messias2014}.}
         \tnt{${^1}$}{A study by \citet{Calanog2014} suggests that sub-millimeter fluxes are magnified by at least a factor of 1.5 greater than the near-IR factors, so we make this assumption here}
\end{deluxetable}

%%%%% MAGPHYS and SEDs %%%%%	
\section{MAGPHYS and Spectral Energy Distributions}
\label{sec:sed}
	The IRAC photometry measured in the 3.6 and 4.5$\mu$m bands, after carefully disentangling the foreground from background components, for the galaxies at $z=1$ to 3, corresponds to rest-frame optical to near-infrared wavelengths. These measurements redwards of the Balmer break allow us to measure the stellar mass and to break certain degeneracies in the SED modeling of DSFGs (though even with excellent photometry, the determination of stellar masses is still degenerate with respect to assumed star-formation histories, see \citealt{Michalowski2012} and \citealt{Michalowski2014}). To model the spectral energy distributions of these DSFGs, we used the Multi-wavelength Analysis of Galaxy Physical Properties ({\sc Magphys}) program developed by \citet{DaCunha2008}. {\sc Magphys} is a model package that empirically derives galaxy properties based on observations at rest wavelengths in the range $912 \rm{\AA} \lesssim \lambda \lesssim 1$mm. 
 %{\sc Magphys} models are calibrated to reproduce the ultraviolet-to-infrared SEDs of local, purely star-forming Ultra Luminous Infrared Galaxies (ULIRGs). 
	
	The SED models assume a Chabrier (2003) initial mass function (IMF) that is cutoff below 0.1 and above 100M$_{\odot}$. If a Salpeter IMF is assumed instead, we find that the stellar masses would be a factor of at least 1.5 larger. The SED models do not include any AGN components, which may contribute significantly to the continuum mid-IR emission. By ignoring AGNs, we could be overestimate the
stellar mass of the galaxies by as much as 0.3 dex \citep{DaCunha2015}.
The templates are based on the Bruzal \& Charlot (2003) spectrum library used as an input for the SED models. These spectra assume an underlying continuous star-formation rate history described by a formation age and a time-scale parameter with random bursts superimposed on the continuous model. We used flux density values from 1$\mu$m to 880$\mu$m shown in Table~\ref{tab:fluxes} as data to be modeled by {\sc Magphys}. The resulting models selected based on Bayesian approach are able to determine the likelihood functions for optical depth of the dust, temperatures of the cold and warm dust components, stellar mass, star-formation rate averaged over the last 100 Myr, and dust mass and luminosity. 

%\input{./Tables/fluxes}
%Table of name, redshift, and fluxes in all wavelengths
\begin{deluxetable*}{l c c c c c c}
        \tablecolumns{7}
        \tablewidth{\textwidth}
        \tablecaption{Intrinsic Flux Densities of SMGs}
        \tablehead{
                \colhead{} &
                \colhead{HBo\"{o}tes02} &
                \colhead{HFLS08} &
                \colhead{G09v1.40} &
                \colhead{G12v2.30} &
                \colhead{G15v2.19} &
                \colhead{HLock12}
        }

        \startdata
        \vspace{0.1cm}
        
        $F_{\rm{F105W}}$ ($\mu$Jy) & --- & --- & --- & --- & $1.63 \pm 0.9$ & ---
        \\ \vspace{0.1cm}
        
        $F_{\rm{F110W}}$ ($\mu$Jy) & --- & $0.7 \pm 0.2$ & --- & --- & $2.6 \pm 0.3$\tnm{${^2}$} & $3.5 \pm 0.7$
        \\ \vspace{0.1cm}
        
        $F_{\rm{J}}$ ($\mu$Jy) & --- & --- & --- & $0.10 \pm 0.02\tnm{${^1}$}$ & $3.4 \pm 0.5$\tnm{${^2}$} & ---
        \\ \vspace{0.1cm}
        
        $F_{\rm{F160W}}$ ($\mu$Jy) & --- & --- & --- & --- & $11.0 \pm 1.1$ & --- 
        \\ \vspace{0.1cm}
        
        $F_{\rm{H}}$ ($\mu$Jy) & --- & --- & --- & --- & $14.2 \pm 2.5$ & ---
        \\ \vspace{0.1cm}
        
        $F_{\rm{K_s}}$ ($\mu$Jy) & $2.5 \pm 1.4$ & --- & $1.31 \pm 0.22$ & $0.74 \pm 0.05\tnm{${^1}$}$ & $12.4 \pm 1.7$ & ---
        \\ \vspace{0.1cm}
        
        $F_{\rm{3.6 {\mu}m}}$ ($\mu$Jy) & $9.0 \pm 2.4$ & $12.2 \pm 2.5$ & $3.7 \pm 0.3$ & $5.0 \pm 0.4$ & $77.9 \pm 6.2$ & $20.9 \pm 2.1$
        \\ \vspace{0.1cm}
        
        $F_{\rm{4.5 {\mu}m}}$ ($\mu$Jy) & $18.1 \pm 4.8$ & $12.6 \pm 2.6$ & $4.2 \pm 0.4$ & $9.0 \pm 0.8$ & $63.9 \pm 5.0$ & $37.7 \pm 3.8$
        \\ \vspace{0.1cm}
        
        $F_{\rm{S250}}$ (mJy)\tnm{${^3}$} & $15.4 \pm 1.0$ & $57.3 \pm 6.67$ & $25.4 \pm 0.7$ & $38.0 \pm 1.3$ & $64.4 \pm 4.6\tnm{${^2}$}$ & $149 \pm 6.7$
        \\ \vspace{0.1cm}
        
        $F_{\rm{S350}}$ (mJy)\tnm{${^3}$} & $18.9 \pm 1.0$ & $62.0 \pm 6.67$ & $24.9 \pm 0.7$ & $46.8 \pm 1.3$ & $34.0 \pm 2.7\tnm{${^2}$}$ & $106 \pm 6.7$
        \\ \vspace{0.1cm}
        
        $F_{\rm{S500}}$ (mJy)\tnm{${^3}$} & $15.1 \pm 1.0$ & $44.7 \pm 6.67$ & $15.8 \pm 0.7$ & $38.8 \pm 1.3$ & $14.7 \pm 1.3\tnm{${^2}$}$ & $52.7 \pm 6.7$
        \\ \vspace{0.1cm}
 
        $F_{\rm{S880}}$ (mJy) & $4.11 \pm 0.82$\tnm{$^4$} & --- & $4.01 \pm 0.94$\tnm{$^4$} & $11.3 \pm 2.3$\tnm{$^4$} & --- & ---

\enddata
  
\tablecomments{De-magnified fluxes and photometric errors, which includes magnification errors. These values were used as inputs for {\sc Magphys} model fitting. Flux densities and errors in $F110W$, $H$, and $K_{\rm s}$ bands are from \citet{Calanog2014} unless otherwise noted. Redshift measurements were obtained using CO observations.}

\tnt{$^1$}{Apparent flux densities and errors from \citet{Fu2012}.}
\tnt{$^2$}{Flux densities and errors from \citet{Messias2014}.}
\tnt{$^3$}{Typical errors in SPIRE photometry are about 10 mJy (magnified), which includes both statistical and confusion noise \citep{Smith2012}. At far-IR wavelengths we assume the foreground lenses are non-dusty galaxies, so the fluxes are due to the background sources only (see \citealt{Wardlow2013}). The exception is G15v2.19, which is lensed by an edge-on, star-forming spiral galaxy.}
\tnt{$^4$}{Flux densities and errors from \citet{Bussmann2013}. Values presented do not include absolute flux density calibration uncertainties of 7\%.}  

  \label{tab:fluxes}
\end{deluxetable*}

%\input{./Tables/properties} 
%Table of Magphys results
\begin{deluxetable*}{l c c c c c c c c c}
\tablecolumns{10}
\tablewidth{\textwidth}
\tablecaption{{\sc Magphys} Results of SMGs}
\tablehead{
        \colhead{Name} &
        \colhead{$f_{\mu}(\rm SFH/IR)\tnm{${^1}$}$} &
        \colhead{${\tau}_{V}$} &
        \colhead{$T^{\rm ISM}_{\rm C}$} &
        \colhead{$T^{\rm BC}_{\rm W}$} &
        \colhead{$M^{*}$} &
        \colhead{SFR} &
        \colhead{sSFR} &%\tnm{${^2}$}$} &
        \colhead{$L_{\rm dust}$} &
        \colhead{$M_{\rm dust}$} 
        \\
        \colhead{} &
        \colhead{} &
        \colhead{} &
        \colhead{[K]} &
        \colhead{[K]} &
        \colhead{[$10^{11} \rm M_{\odot}$]} &
        \colhead{[$\rm M_{\odot} yr^{-1}$]} &
        \colhead{[$10^{-10}$yr$^{-1}$]} &
        \colhead{[$10^{11} \rm L_{\odot}$]} &
        \colhead{[$10^{8} \rm M_{\odot}$]}
}
\startdata

%%%%% HBootes02 %%%%%
\vspace{-0.2cm} 
& $0.51^{+0.09}_{-0.11}$ 
\\
HBo\"{o}tes02 & & $4.81^{+0.71}_{-0.69}$ & $24.6^{+1.6}_{-3.3}$ & $41.7^{+8.0}_{-7.5}$ & $4\pm1$ & $383^{+12}_{-13}$ & $9.3\pm1.0$ & $74.2^{+0.8}_{-0.9}$ & $15.2^{+2.5}_{-1.4}$
\\
\vspace{0.2cm} 
& $0.37^{+0.12}_{-0.11}$
\\

%%%%% HFLS08 %%%%%
\vspace{-0.2cm}
& $0.34^{+0.07}_{-0.18}$ 
\\
HFLS08 & & $4.12^{+1.08}_{-0.40}$ & $24.9^{+2.8}_{-3.4}$ & $45.1^{+8.3}_{-5.6}$ & $0.9 \pm 0.1$ & $412^{+10}_{-40}$ & $49\pm1.5$ & $137^{+1.0}_{-1.0}$ & $34.8^{+1.4}_{-0.8}$
\\
\vspace{0.2cm}
& $0.29^{+0.13}_{-0.13}$
\\

%%%%% G09v1.40 %%%%%
\vspace{-0.2cm}
& $0.33^{+0.08}_{-0.16}$
\\
G09v1.40 & & $5.19^{+1.07}_{-0.40}$ & $24.6^{+2.1}_{-3.5}$ & $41.7^{+9.3}_{-6.1}$ & $0.8\pm0.1$ & $129^{+34}_{-8}$ & $16.5 \pm 1.0$ & $51.7^{+2.2}_{-2.4}$ & $8.99^{+1.2}_{-1.2}$
\\
\vspace{0.2cm}
& $0.37^{+0.16}_{-0.16}$
\\

%%%%% G12v2.30 %%%%%
\vspace{-0.2cm}
& $0.66^{+0.00}_{-0.05}$
\\
G12v2.30 & & $3.79^{+0.26}_{-0.03}$ & $25.0^{+1.5}_{-4.9}$ & $53.1^{+4.9}_{-6.7}$ & $2.9\pm0.6$ & $101^{+45}_{-1}$ & $50.7\pm0.6$ & $190^{+1.1}_{-1.0}$ & $42.4^{+0.7}_{-0.6}$
\\
\vspace{0.2cm}
& $0.52^{+0.18}_{-0.05}$
\\

%%%%% G15v2.19 %%%%%
\vspace{-0.2cm} 
& $0.68^{+0.03}_{-0.12}$ 
\\
G15v2.19 & & $4.22^{+0.39}_{-0.19}$ & $24.9^{+0.2}_{-0.2}$ & $37.6^{+7.18}_{-6.50}$ & $1.8\pm0.2$ & $98.8^{+29}_{-5}$ & $5.3\pm0.5$ & $23.7^{+0.5}_{-0.4}$ & $7.75^{+0.25}_{-0.25}$
\\
\vspace{0.2cm} 
& $0.53^{+0.11}_{-0.15}$
\\

%%%%% HLock12 %%%%%
\vspace{-0.2cm}& $0.33^{+0.14}_{-0.00}$ 
\\
HLock12 & & $5.19^{+0.84}_{-0.00}$ & $25.0^{+1.9}_{-3.6}$ & $44.8^{+4.5}_{-3.2}$ & $3.0\pm0.4$ & $495^{+37}_{-4}$ & $16.5\pm0.5$ & $202^{+1.1}_{-1.0}$ & $29.7^{+0.7}_{-0.7}$
\\
\vspace{0.2cm}
& $0.20^{+0.13}_{-0.05}$

\enddata
\tablecomments{$f_{\mu}$ is the fractional energy absorbed by the ISM, calculated from stellar-dominated (SFH) and dust-dominated (IR) photometry. ${\tau}_{V}$ is the total $V$-band optical depth of the dust seen by young stars in their birth clouds. $T^{\rm ISM}_{\rm C}$ is the equilibrium temperature of cold dust in the ambient ISM. $T^{\rm BC}_{\rm W}$ is the equilibrium temperature of warm dust in stellar birth clouds. $M^{\star}$ is the stellar mass. SFR is the star-formation rate averaged over $10^{8}$ years, while ${\psi}_{S} = \rm{SFR} / M^{\star}$ is the specific star-formation rate normalized to stellar mass. $L_{\rm dust}$ and $M_{\rm dust}$ are the total dust luminosity and mass, respectively. Reported values are the best-fit values with their 16th and 84th percentiles.}
\tnt{${^1}$}{Values in the first and second rows give $f_{\mu}^{\rm SFH}$ and $f_{\mu}^{\rm IR}$, respectively.}
%\tnt{${^2}$}{Averaged SFR over $10^{8}$ years.}

\label{tab:properties}
\end{deluxetable*}
		
%%%%% Results and Discussion %%%%%
%\clearpage
\section{Results and Discussion}
\label{sec:results}

	A full summary of the fitted parameters for the six lensed DSFGs is found in Table~\ref{tab:properties}. We also show the best-fit total and unattenuated spectral energy distributions for each of the DSFGs in Figure~\ref{fig:sed}.

	Since rest-frame absolute \textit{H}-band magnitudes (\textit{M}$_{\rm{H}}$) provide a guide to galaxy stellar masses that is not complicated by details of the assumed star-formation history, we compare the predicted values based on the best-fit SED models to other samples in Figure 2. The data comes from 850 and 870$\mu$m-selected SMGs \citep{Hainline2011,Wardlow2011}. The {\it Herschel}-selected lensed galaxies are on average consistent with the rest-frame H-band magnitude distribution of 850 and 870$\mu$m-selected SMGs. Using a standard Kolmogorov-Smirnov test on the two populations of SMGs identified from our {\it Herschel} observations and that of 850 and 870$\mu$m observations, we find a K-S statistics of 0.253 and a probability value of 0.648, suggesting that the two data sets are from the same parent population (null hypothesis). Thus if there is no active galactic nucleus contribution to their H-band luminosities, and the star-formation histories are similar, they are likely to have stellar masses consistent with other starbursting galaxy samples.

	In Figure 3 we compare stellar mass, IR luminosity, and gas mass fraction of the sample to other {\it Herschel}-selected lensed galaxies from the literature, following \citet{Negrello2014}. Among the six DSFGs presented here, we find gas mass values from CO molecular line observations for two of the galaxies (G12v2.30 from \citealt{Fu2012} and G15v2.19 from \citealt{Messias2014}). We make use of the stellar mass estimates from {\sc Magphys} modeling to calculate $f_{\rm gas}=M_{\rm gas}/(M_{\rm gas}+M^*)$ and find values of $0.20 \pm 0.06$ and $0.20 \pm 0.07$ for G12v2.30  and G15v2.19, respectively. These are comparable to the gas mass fraction values quoted in the literature.

	We find that a significant attenuation by dust ($\tau_V \approx 4$--5) is required to be consistent with the {\it HST}/WFC3, Keck/NIRC2, and IRAC photometry. Such high extinction values are consistent with other ULIRGs and SMGs (e.g. \citealt{Geach2007,DaCunha2010,Michalowski2010,Hainline2011,Wardlow2011,Rowlands2014}). The stellar masses of these galaxies span the range of $8\times10^{10}$ -- $4\times10^{11}$ M$_{\odot}$ while their 100 Myr-averaged star-formation rates range from 100 to 500 M$_{\odot}$ yr$^{-1}$. These star-formation rates are consistent with the starburst nature of these DSFGs, some of which can also be classified as SMGs based on their lensing magnification-corrected sub-millimeter flux densities.

	The physical properties that we derive for our sample of lensed DSFGs clearly indicate fundamental differences between this population and other star-forming galaxies identified from optical surveys at similar or lower redshifts. In Figure~\ref{fig:smg} we show the SFR (inferred from the SED) vs stellar masses for these galaxies and compare our sample of DSFGs to star-forming galaxies and SMGs from the literature. For reference we also plot the average ``main sequence'' relations for galaxies at $z =1$ and 2. The horizontal lines show the selection, given that the magnification-corrected 500$\mu$m flux density values have a minimum at the level of 10mJy for our sample. The selection suggests the redshift dependence of the star-formation rate vs stellar mass for our sample and other SMGs, since the selection based on a flux density results in selecting starbursts with higher star-formation rates with increasing redshift.

	Systems identified from the sub-millimeter surveys as SMGs have relatively large estimated stellar masses compared to the normal star-forming galaxies identified from standard surveys \citep{Magnelli2012}. In fact, our lensed {\it Herschel}-selected DSFGs have SED-inferred stellar mass estimates in the range of $\sim 8 \times 10^{10} - 4 \times 10^{11}$ M$_{\odot}$ with a mean value of $\sim 2.3 \times 10^{11}$ M$_{\odot}$. This tight range of stellar masses is believed to not be related to the optical and near-IR sensitivities for these systems, as lower mass galaxies are expected to be detected in the deep Spitzer observations \citep{Reddy2006, Michalowski2010}. As mentioned earlier, we do not account for an AGN contribution to the SED of these galaxies. The current data do not allow us to separate AGN templates from SF in galaxies, while existing studies that separated AGN activity relied on significant mid-IR coverage, including {\it Spitzer}/IRS spectra \citep{Berta2013}. As discussed in \citet{DaCunha2015}, stellar mass is the single parameter that is most affected by an AGN contribution to the SED that is ignored in model fits, and we may have over-estimated the stellar mass by 0.3 dex. The star-formation rates measured for SMGs contribute very little to the stellar mass, with most of the mass being built up before the sub-millimeter phase \citep{Michalowski2010}. This explains the scatter that we see in the M$^{\star}$-SFR plane for the SMG population, with rates ranging from $\sim$ 100 - 400 M$_{\odot}$ yr$^{-1}$. 

While our lensed DSFGs have high stellar masses, comparable to SMGs, their sSFRs may not amount to the high level seen in SMGs with values above 5 Gyr$^{-1}$. 	Star-forming galaxies at $z\sim2$ on the main sequence of that redshift have sSFRs at the level of 2 Gyr$^{-1}$ or below \citep{Fu2013}. Based on SED models, only two of our targets have sSFRs that are at the level of 5 Gyr$^{-1}$ (G12v2.30 and HFLS08). It is yet unclear if all SMGs are a result of galaxy mergers, however, a previous multi-wavelength study of G12v2.30 \citep{Fu2012} and the resulting lensing model did find clear evidence for a merger in the source plane of that system. Unfortunately, HFLS08 does not have the same level of multi-wavelength data as G12v2.30 and the lens reconstruction \citep{Calanog2014} for near-IR observations only shows a single galaxy. This is somewhat misleading since other lensed SMGs/DSFGs, including G12v2.30, show multi-component structure in the source plane with rest-frame optical emission spatially disjoint from the peak mm-wave dust emission where the starburst is active (e.g., \citealt{Fu2012,Dye2015,Hodge2015}). 

The fact that only galaxies with sSFRs greater than 5 Gyr$^{-1}$  are SMGs does not imply that DSFGs with sSFRs greater than 5 Gyr$^{-1}$ are single galaxies. G15v2.19 with a sSFR of roughly 1 Gyr$^{-1}$ shows a clear merger with again distinctively separate peak rest-frame optical and dust emission \cite{Messias2014,Timmons2015}. It may be that a majority of the lensed DSFGs are the result of merger-driven star formation activities. Higher resolution observations are clearly desirable to address this, not just in the millimeter wavelengths with interferometers, but also in the optical and infrared. While Keck in the $K_{\rm s}$ band and HST/WFC3 at 1.6$\mu$m and below can provide the necessary data to study DSFGs in high resolution, our study finds that observations with {\it Spitzer}/IRAC are also desirable to determine key parameters from the SED, especially stellar mass and galaxy extinction. Such information can be extracted from the data despite issues related to the low spatial resolution in the data involving lensed systems. In the era of JWST we may be able to do more since imaging then at wavelengths around 3-4$\mu$m will provide necessary spatial resolution at the same level as interferometers. In that case we may not be able to just obtain globally averaged values, but also perform SED modeling of the clumpy interstellar medium of lensed DSFGs at spatial resolution of 300pc, as achieved by mm-wave interferometer data such as ALMA \citep{Dye2015} and establish the stellar masses of individual clumps.

\begin{figure}
	\begin{minipage}{\linewidth}
		\includegraphics[scale=0.55,trim = 1.5cm 0 0 0,clip]{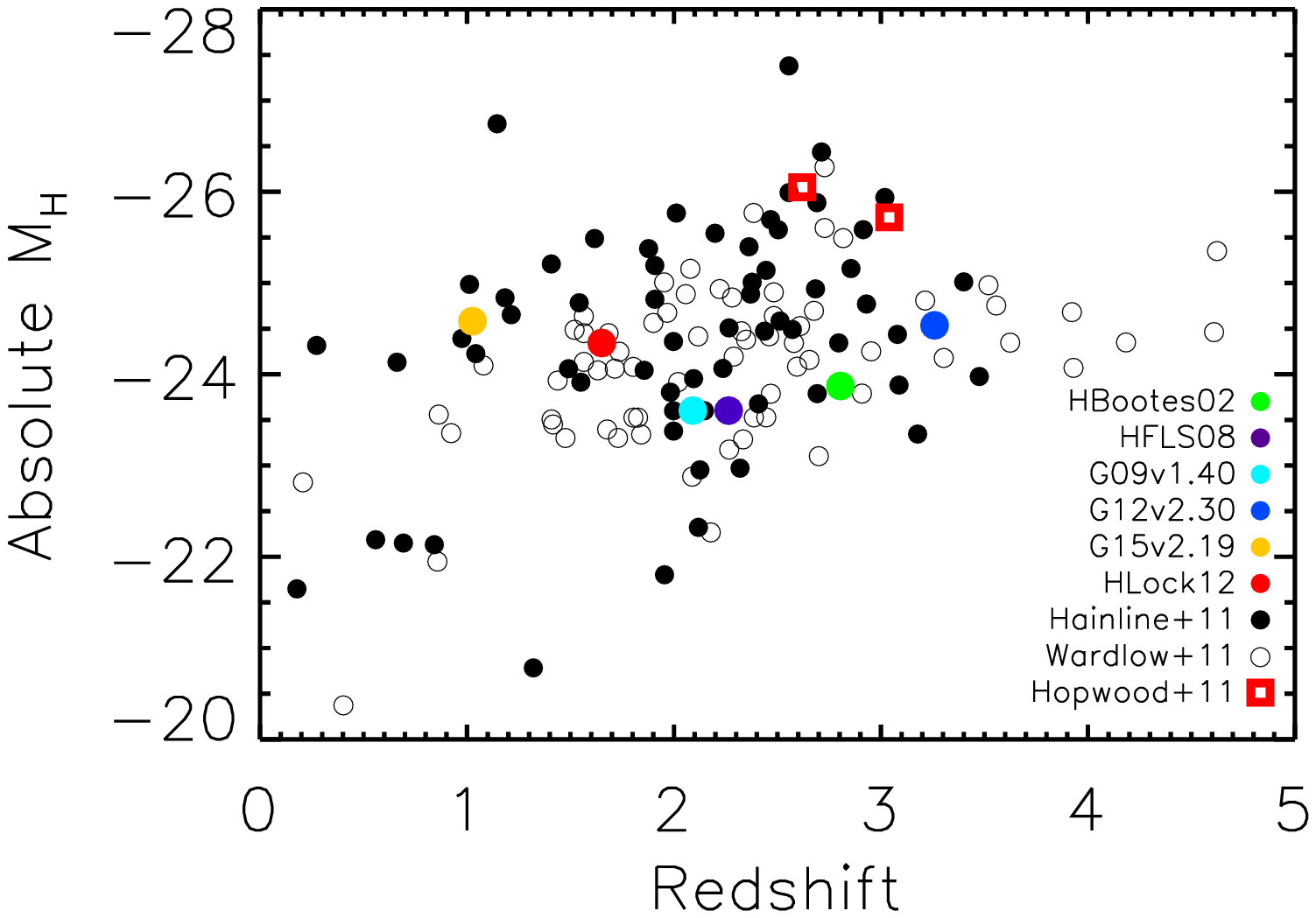}
		\caption{Magnification-corrected, rest-frame absolute $H$-band magnitude $M_{\rm H}$ versus redshift for DSFGs. We also include 850 and 870${\mu}$m-selected galaxies \citep{Hainline2011,Wardlow2011}, as well as {\it Herschel}-selected targets SDP.81 and SDP.130 \citep{Hopwood10}. A Kolmogorov-Smirnov statistic of 0.253 with a significance level of 0.648 suggests that {\it Herschel}-selected lensed galaxies are consistent with the rest-frame H-band magnitude distribution of 850 and 870$\mu$m-selected SMGs.}
		\label{fig:hmag}
	\end{minipage}
\end{figure}

\begin{figure}
	\begin{minipage}{\linewidth}
		\hspace{-2.5cm}
		\includegraphics[scale=0.7]{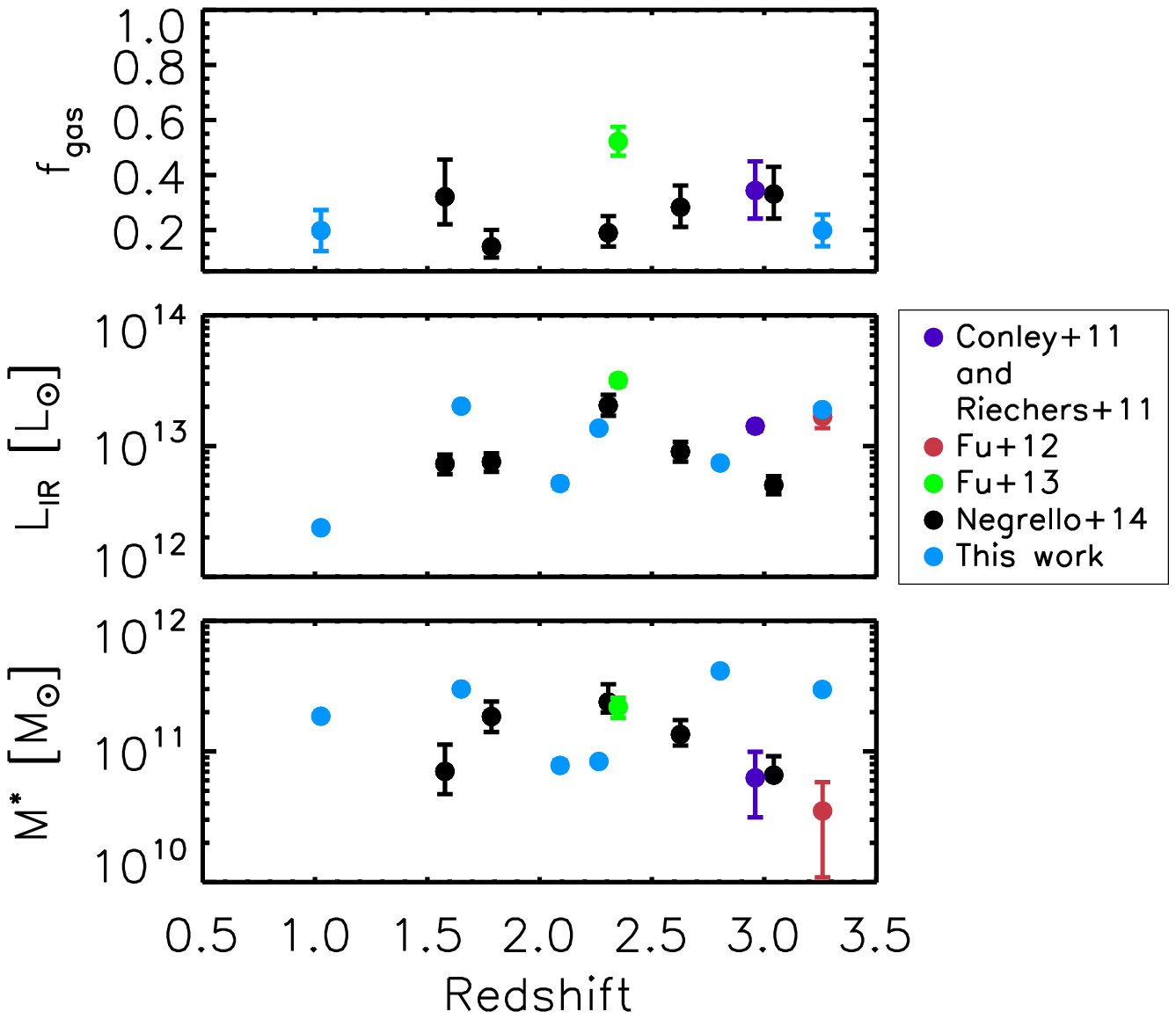}
%		\vspace{0.2cm}
		\caption{Top: gas fraction $f_{\rm gas} = M_{\rm gas} / (M_{\rm gas} + M_{\rm stellar})$ vs. redshift. We use gas mass values presented by \citet{Fu2012} and \citet{Messias2014} to determine gas fractions for G12v2.30 and G15v2.19, respectively. Gas fraction value for the purple data point is provided by \citet{Riechers2011}. Middle: magnification-corrected infrared luminosity $L_{\rm IR}$ vs. redshift. Here we assume dust luminosities dominate the contribution to IR luminosities for our sources, as indicated by the blue points on the plot. Infrared luminosity value for the purple data point is provided by \citet{Conley2011}. Bottom: magnification-corrected stellar mass $M^*$ (bottom) vs. redshift. Included are SMGs---lensed and unlensed---examined in other studies: \citet{Conley2011,Fu2012,Fu2013}; and \citet{Negrello2014}. Stellar mass value for the purple data point is provided by \citet{Conley2011}.}
		\label{fig:comp}
	\end{minipage}
\end{figure}

%\vspace{2cm}

\begin{figure*}
	\begin{center}
		\hspace{-1cm}
		\includegraphics[scale=0.85]{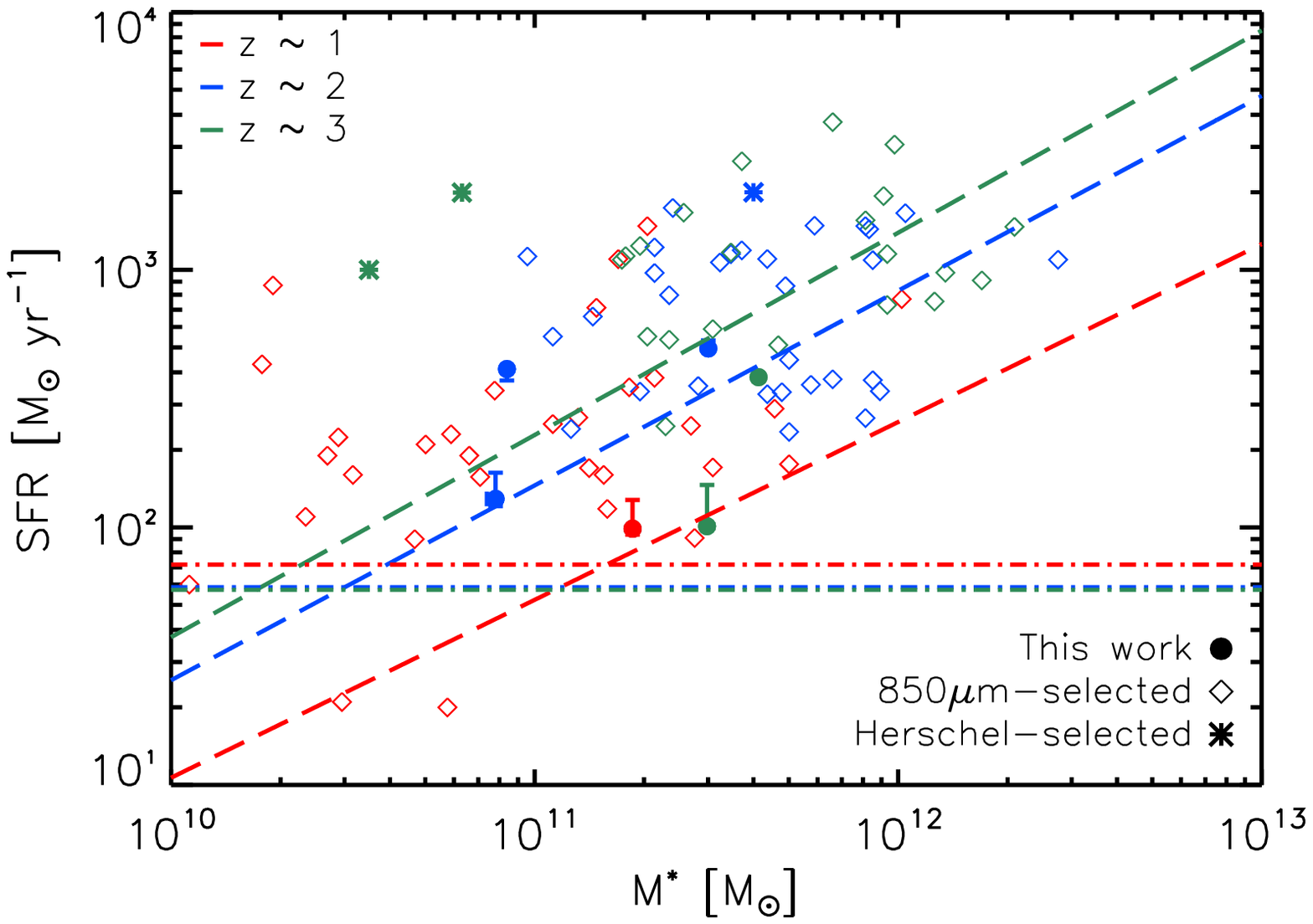}
		\caption{Star-formation rate versus stellar mass of SMGs and DSFGs. Circles represent sources in this study, and we also include data from \citet{Michalowski2010}, \citet{Banerji2011}. \citet{Conley2011}, \citet{Fu2012}, and \citet{Fu2013}. 850$\mu$m-selected SMGs are shown as diamonds and {\it Herschel}-selected targets are shown as stars. Red points correspond to $z \sim 1$, blue points $z \sim 2$, and green points $z \sim 3$. The dashed lines show the main sequence correlations at each redshift \citep{Speagle2014}. The horizontal dash-dotted lines represent the $S_{500} = 10$ mJy selection cutoff for SMGs, drawn from mock SEDs at each redshift.}
		\label{fig:smg}
	\end{center}
\end{figure*}

%\vspace{2cm}

\section{Summary}

	We have presented the rest-frame optical spectral energy distribution and physical properties of six {\it Herschel}-selected gravitationally lensed dusty, star-forming galaxies (DSFGs) at redshifts of 1 to 3. The lensed DSFGs were first identified with {\it Herschel}/SPIRE imaging data from {\it Herschel} Astrophysical Terahertz Large Area Survey (H-ATLAS) and {\it Herschel} Multi-tiered Extragalactic Survey (HerMES). 
	
\begin{enumerate}
	\item	The targets were observed with {\it Spitzer}/IRAC at 3.6 and 4.5$\mu$m. Due to the spatial resolution of the IRAC observations at the level of 2$''$, the lensing features of the background DSFG is blended with the flux from the foreground lensing galaxy in the IRAC imaging data. We make use of higher resolution {\it Hubble}/WFC3 or Keck/NIRC2 Adaptive Optics imaging data to fit light profiles of foreground lensing galaxies as a way to model the foreground components in order to disentangle the foreground lens and background source flux densities. The flux density measurements at 3.6 and 4.5$\mu$m, once combined with {\it Hubble}/WFC3 and Keck 2.2$\mu$m data, provide important constraints on the rest-frame optical spectral energy distribution of the {\it Herschel}-selected lensed DSFGs. We model the combined UV to mm-wavelength SEDs to establish the stellar mass, dust mass, star-formation rate, and visual extinction, among other parameters for each of these {\it Herschel}-selected DSFGs. 
	
	\item {\it Herschel}-selected lensed galaxies are consistent with the rest-frame H-band magnitude distribution of 850 and 870$\mu$m-selected SMGs. Assuming no contribution from AGN to their H-band luminosities, these galaxies are likely to have stellar masses consistent with other starbursting galaxy samples. {\it Herschel} selections may also depend on redshift, since the selection based on flux densities results in selecting starbursts with higher SFRs with increasing redshift.
	
	\item The high extinction values for these DSFGs ($\tau_{\rm V} \approx$ 4-5) are consistent with other ULIRGs and SMGs. The stellar masses of these galaxies span the range of $8\times10^{10}$ -- $4\times10^{11}$ M$_{\odot}$, while their star-formation rates range from 100 to 500 M$_{\odot}$ yr$^{-1}$.  The inferred SFRs are consistent with the starburst nature of DSFGs, some of which can also be classified as SMGs based on their flux densities.
%	The star-formation rate vs.~stellar mass for the six galaxies fall on average statistically higher than the mean main sequence relations at $z \sim 1$ to 3. We find that such a placement may be due to a selection effect.
		
	\item We find a large scatter between stellar mass and SFR for the SMG population. However, we also observe a correlation between the specific star-formation rate and dust temperature. This suggests that these galaxies were formed by merging systems. 
	
	\item Lensed systems are intrinsically similar to far-infrared selected samples, which allows analyses of lensed galaxies to be generalized to the larger population of SMGs and DSFGs, and vice-versa. 
	
	\item We conclude that, despite typically low resolutions, {\it Spitzer}/IRAC data provide vital constraints on key parameters such as stellar mass. Followup observations with higher resolutions by instruments such as the future JWST can not only introduce further constraints, but may also allow for detailed analyses of the ISM of DSFGs with resolutions of $\sim$ 300pc, as achieved by previous interferometers such as ALMA.
	
\end{enumerate}
	
\section*{acknowledgments}

	Financial support for this work was provided by NASA through the Spitzer Space Telescope, which is operated by the Jet Propulsion Laboratory, California Institute of Technology under a contract with NASA. 

	Additional support for AC, BM, HN, CC, and NT was from the National Science Foundation with AST-1313319.

	Some of the data presented herein were obtained at the W.M. Keck Observatory, which is operated as a scientific partnership among the California Institute of Technology, the University of California and the National Aeronautics and Space Administration. The Observatory was made possible by the generous financial support of the W.M. Keck Foundation. We thank the native communities of Hawaii for their support of Keck and the Mauna Kea observatories. We recognize and acknowledge the very significant cultural role and reverence that the summit of Mauna Kea has always had within the indigenous Hawaiian community.  We are most fortunate to have the opportunity to conduct observations from Mauna Kea. We encourage a meaningful dialogue between the astronomical and native Hawaiian communities on the future of Mauna Kea and the astronomical facilities there.

	The {\it Herschel}-ATLAS is a project with Herschel, which is an ESA space observatory with science instruments provided by European-led Principal Investigator consortia and with important participation from NASA. The H-ATLAS web site is \url{http://www.h-atlas.org/}.

	This research has made use of data from the HerMES project (\url{http://hermes.sussex.ac.uk/}). HerMES is a {\it Herschel} Key Programme utilizing Guaranteed Time from the SPIRE instrument team, ESAC scientists and a mission scientist. The data presented in this paper will be released through the HerMES Database in Marseille, HeDaM (\url{http://hedam.oamp.fr/HerMES/}).

	The Dark Cosmology Centre is funded by the Danish National Research Foundation.

	IO acknowledges support from the European Research Council (ERC) in the form of Advanced Grant, {\sc cosmicism}

%%%%%References%%%%%
\bibliographystyle{apj_hack}
\bibliography{sed_draft}

%%%%% Appendix %%%%%
\newpage

\section{Appendix}
%\vspace{1cm}
\begin{figure}[h]
%	\vspace{1.2cm}
	\begin{minipage}{\linewidth}
		\includegraphics[scale=0.5]{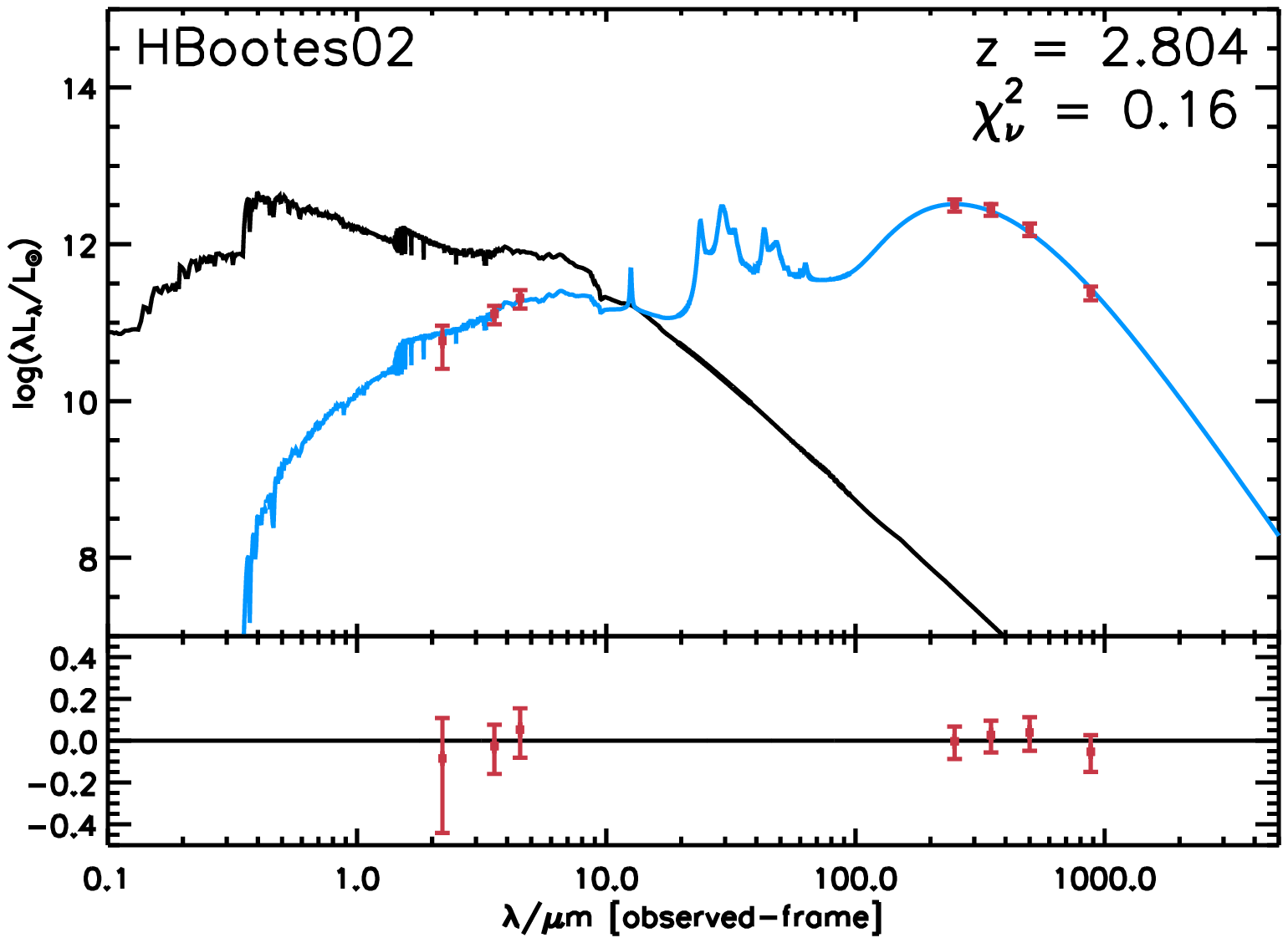} 
		\includegraphics[scale=0.5]{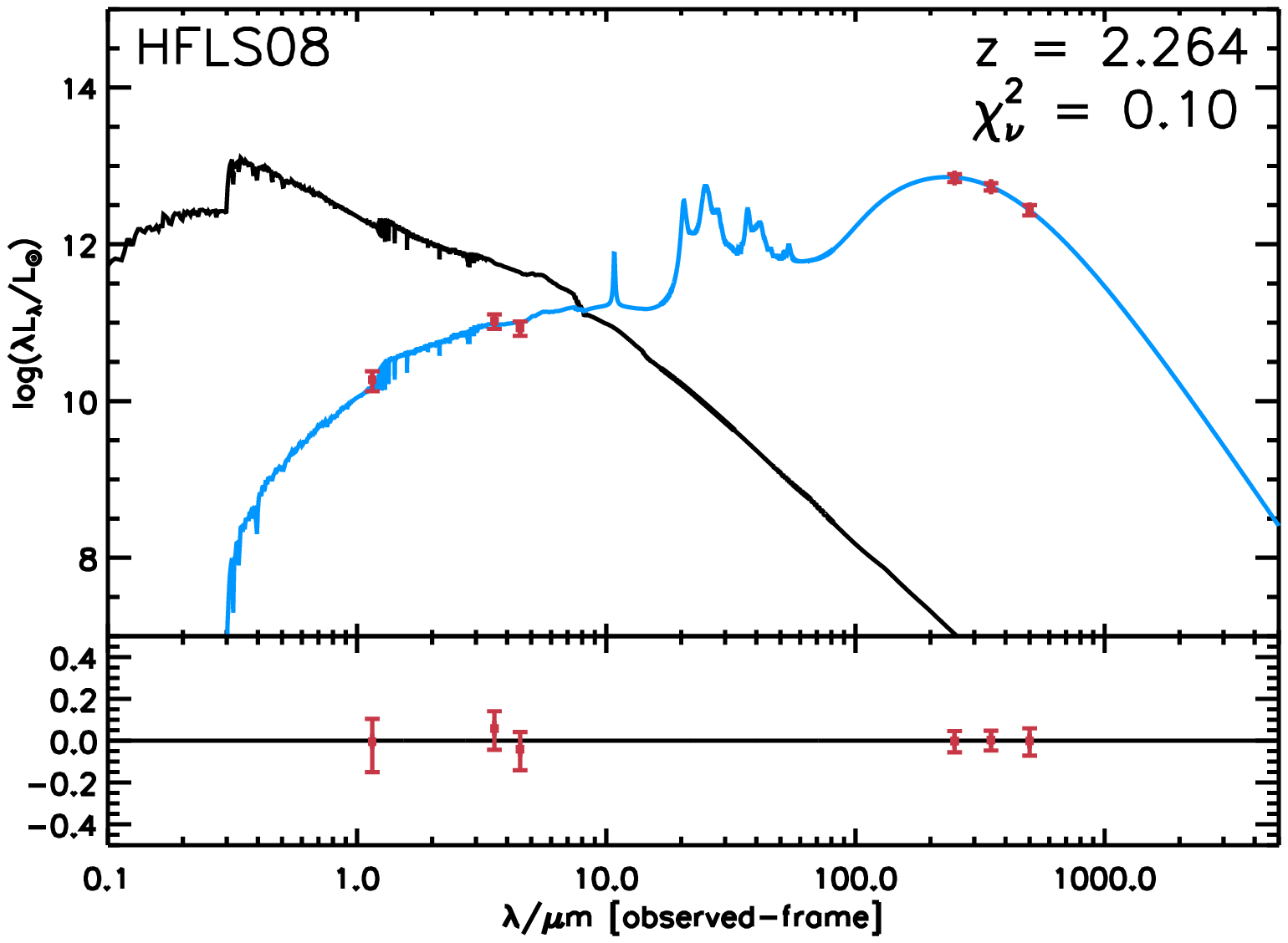} 
		\includegraphics[scale=0.5]{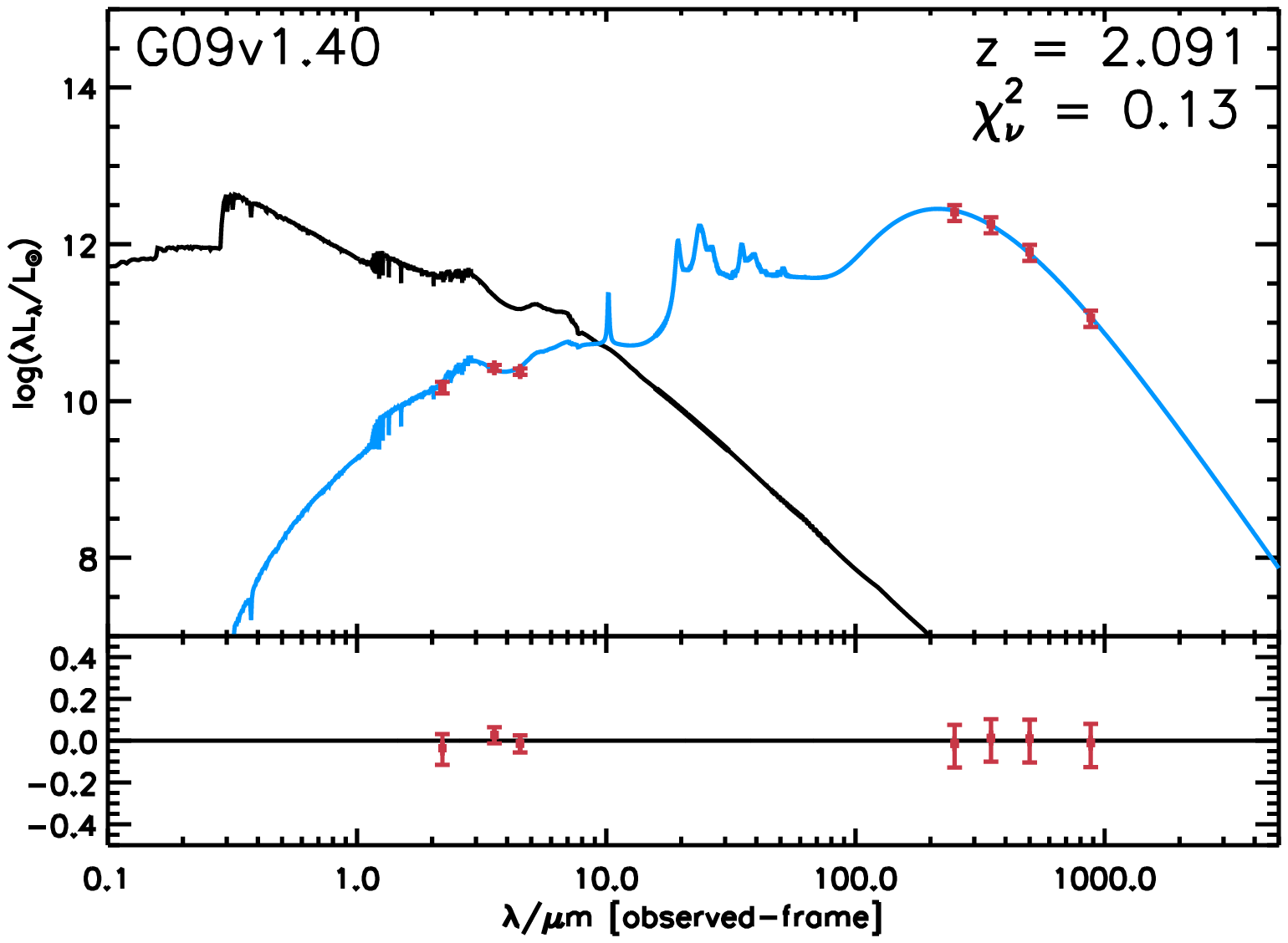} 
		\includegraphics[scale=0.5]{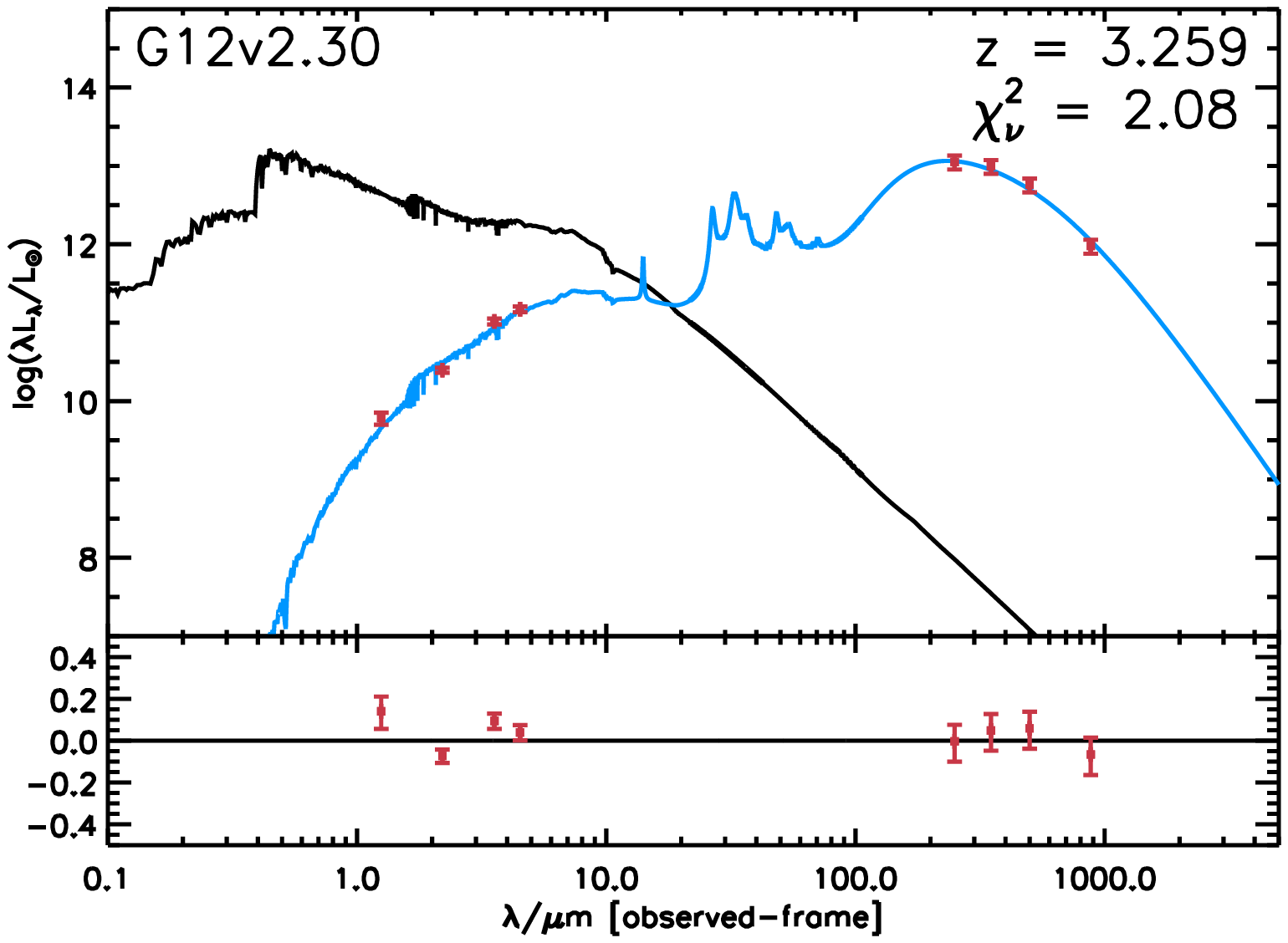} 
		\includegraphics[scale=0.5]{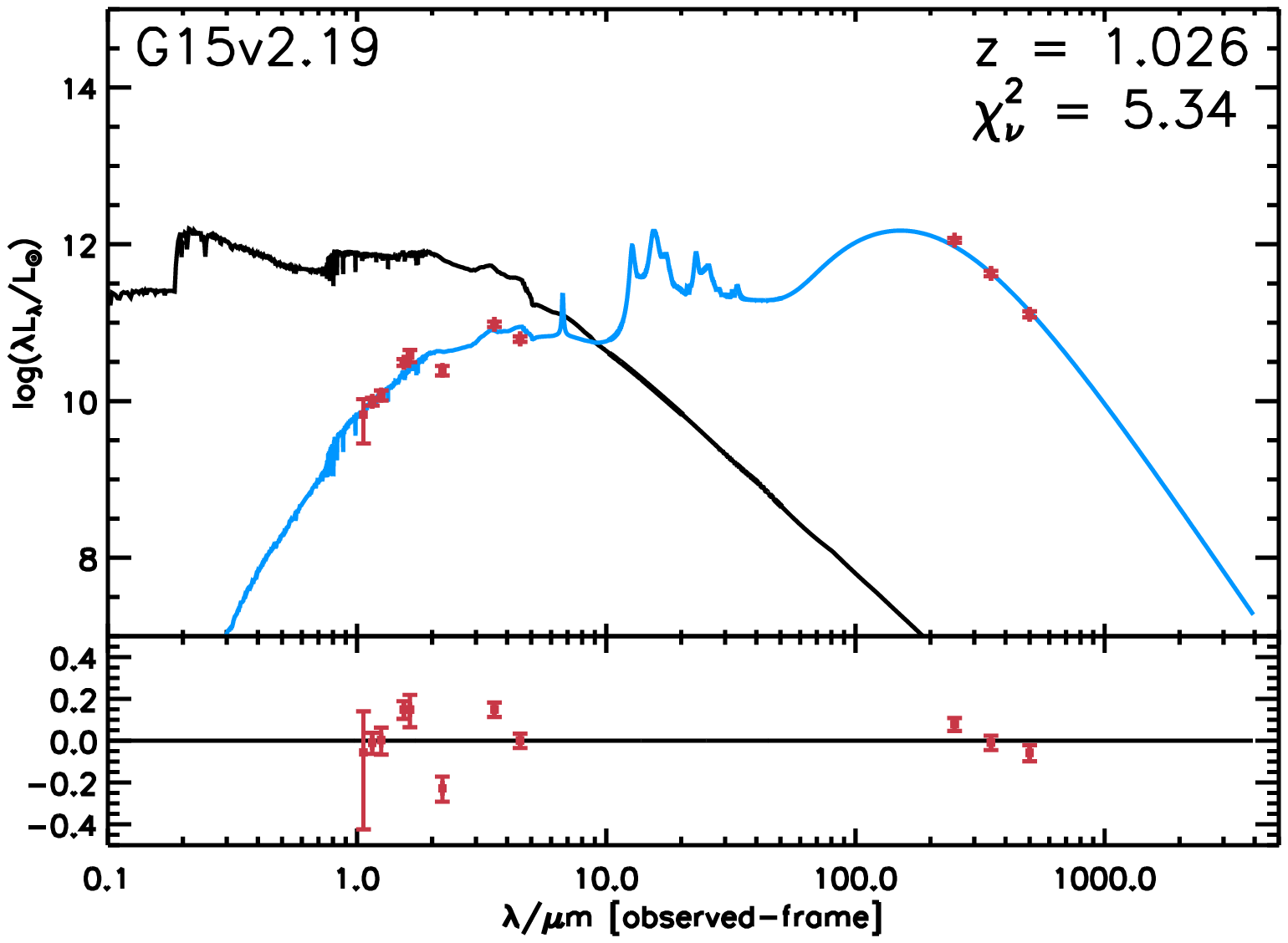} 
		\hspace{0.1cm}
		\includegraphics[scale=0.5]{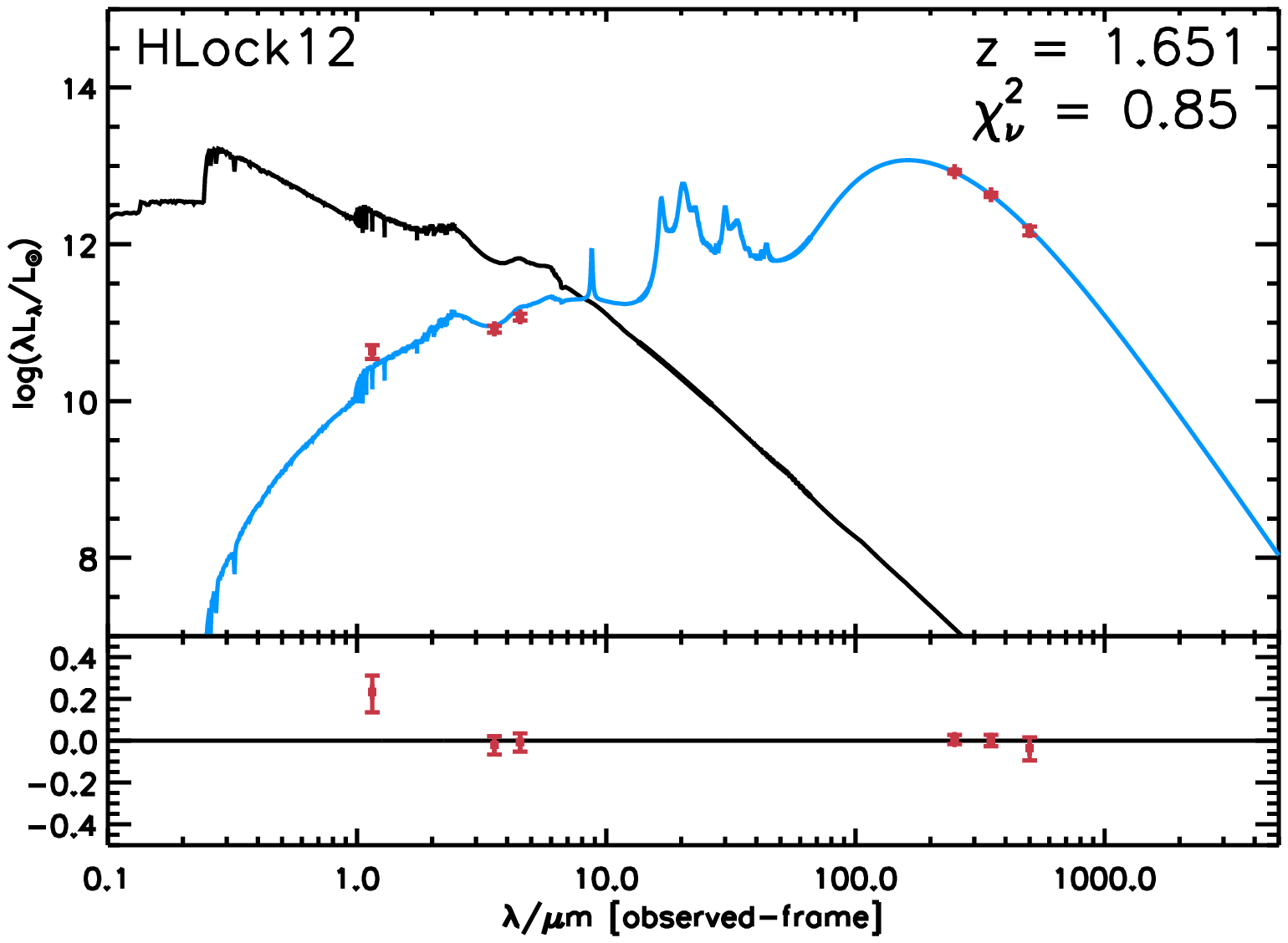} 
	\end{minipage}
	\vspace{1cm}
	\caption{Spectral energy distributions of each DSFG. Reported redshifts correspond to their respective sources. The red data points are the input fluxes (see Table~\ref{tab:fluxes}). The blue and black curves are the best-fit attenuated and unattenuated SED's, repectively. Redshifts correspond to the DSFGs and $\chi^2_\nu$ statistics indicate the goodness of fits for each SED. The lower panels display the residuals for each fit.}
	\label{fig:sed}
\end{figure}

\begin{figure}
	\begin{center}
		\includegraphics[trim = 2cm 1.2cm 0.5cm 3.8cm,clip]{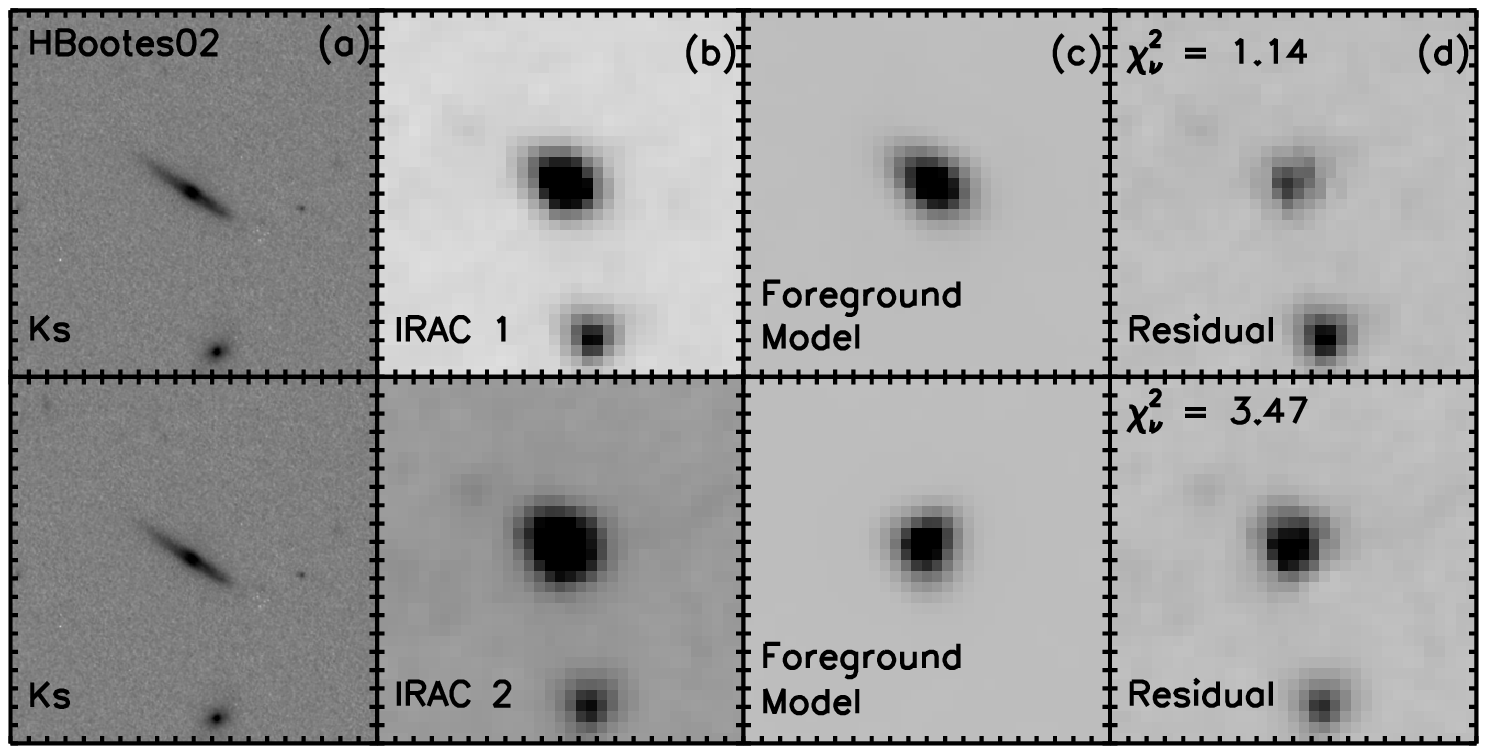}
		\includegraphics[trim = 2cm 1.2cm 0.5cm 3.8cm,clip]{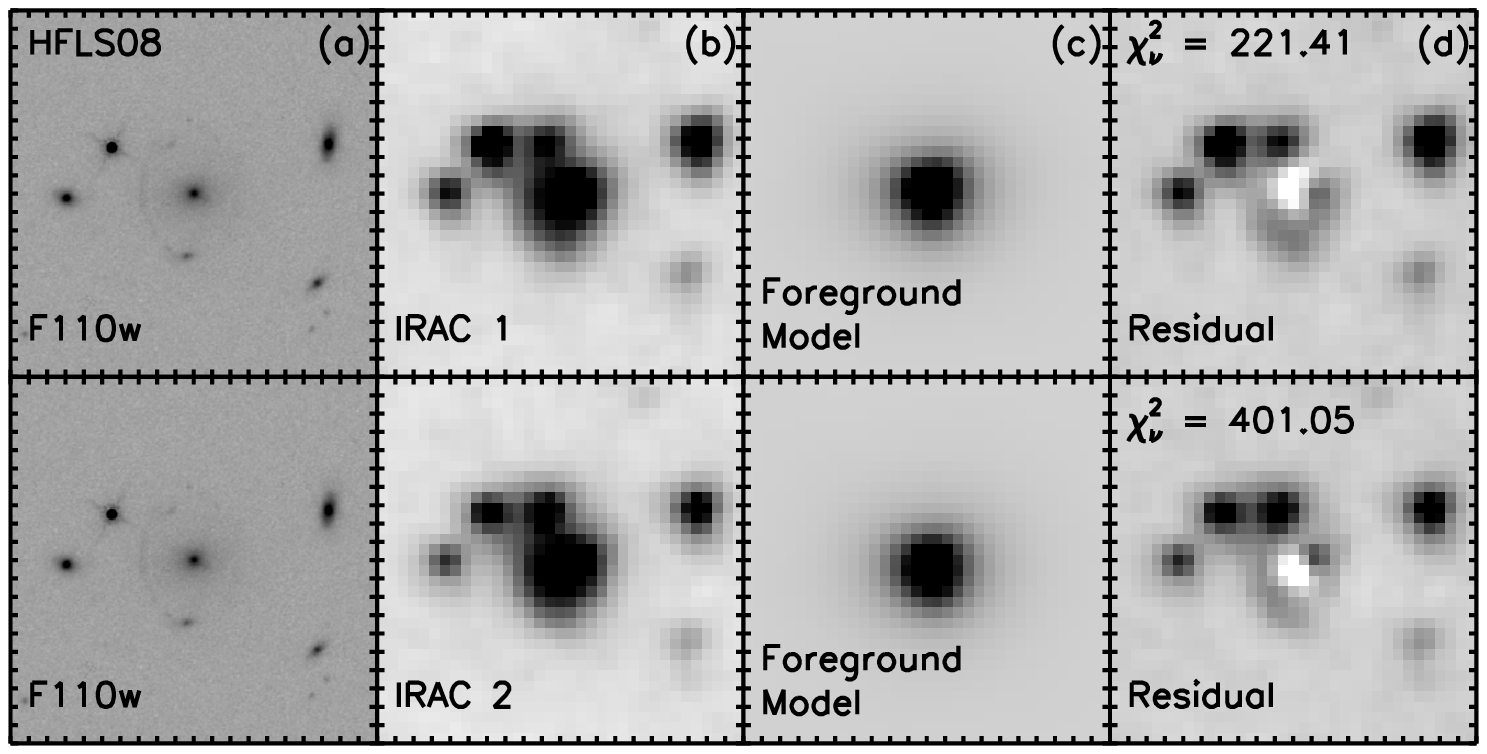} 
		\includegraphics[trim = 2cm 1.2cm 0.5cm 3.8cm,clip]{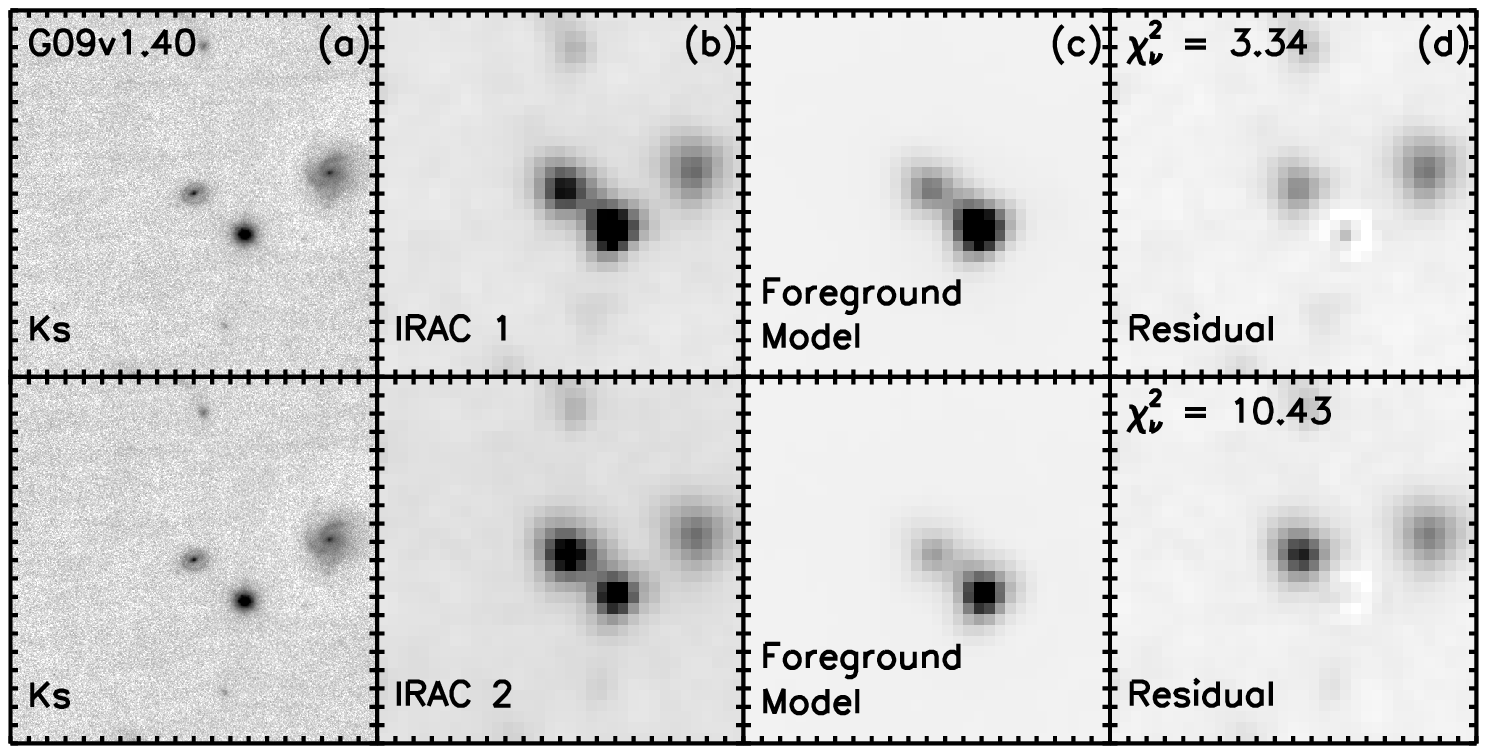}
	\end{center}
	\caption{{\it Spitzer} IRAC de-blending results. Upper rows correspond to IRAC Channel 1 (3.6$\mu$m) and lower rows correspond to IRAC Channel 2 (4.5$\mu$m). From left to right we show: (a) high-resolution data, (b) the original {\it Spitzer} IRAC data, (c) the best-fit foreground model, and (d) the residual image with the foreground component(s) removed. We also show the $\chi^2_\nu$ statistic for each fit. Images are oriented such that North is up and East is left. Each cutout has size $20'' \times 20''$, with each tickmark representing 1$''$.}
	\label{fig:deblend}
\end{figure}
\begin{figure}
	\begin{center}
		\includegraphics[trim = 2cm 1.2cm 0.5cm 3.8cm,clip]{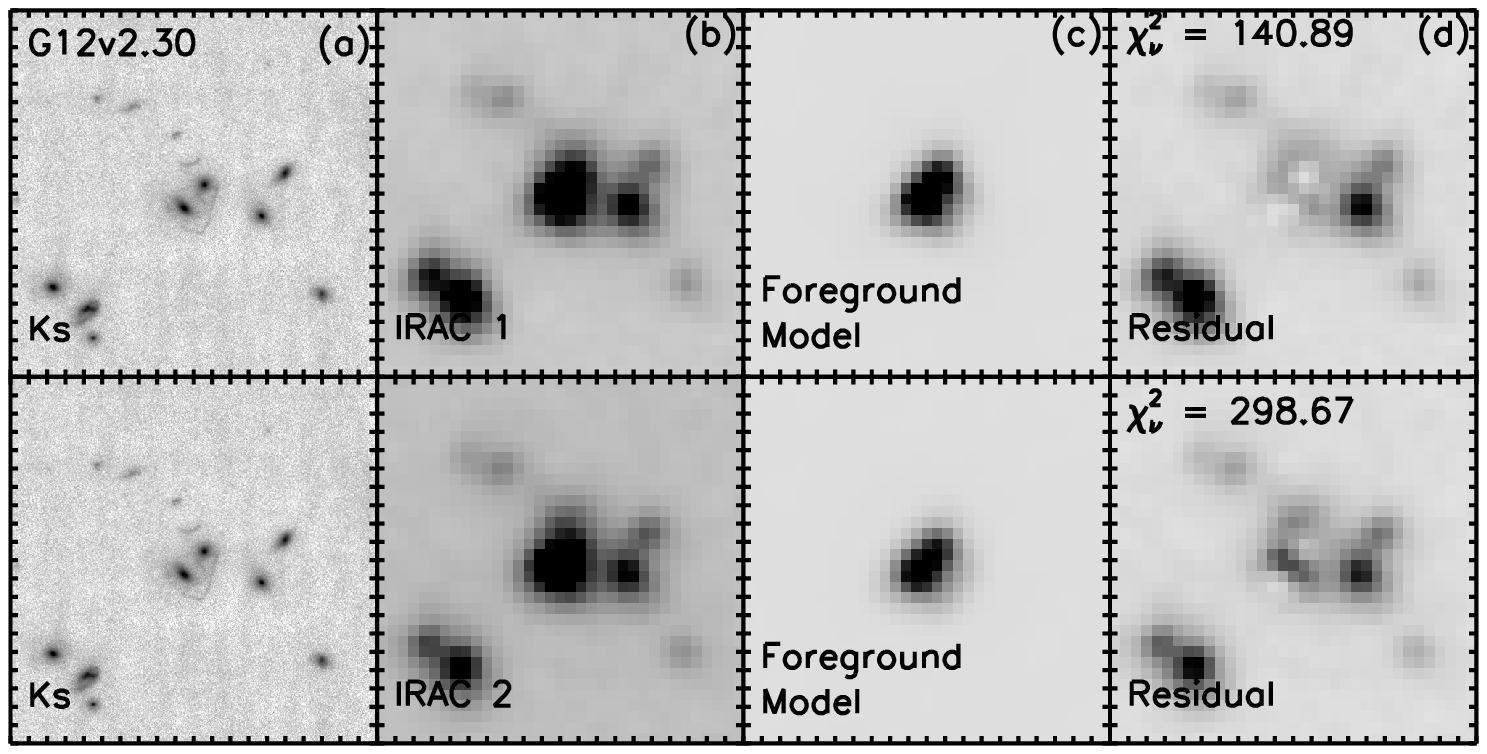} 
		\includegraphics[trim = 2cm 1.2cm 0.5cm 3.8cm,clip]{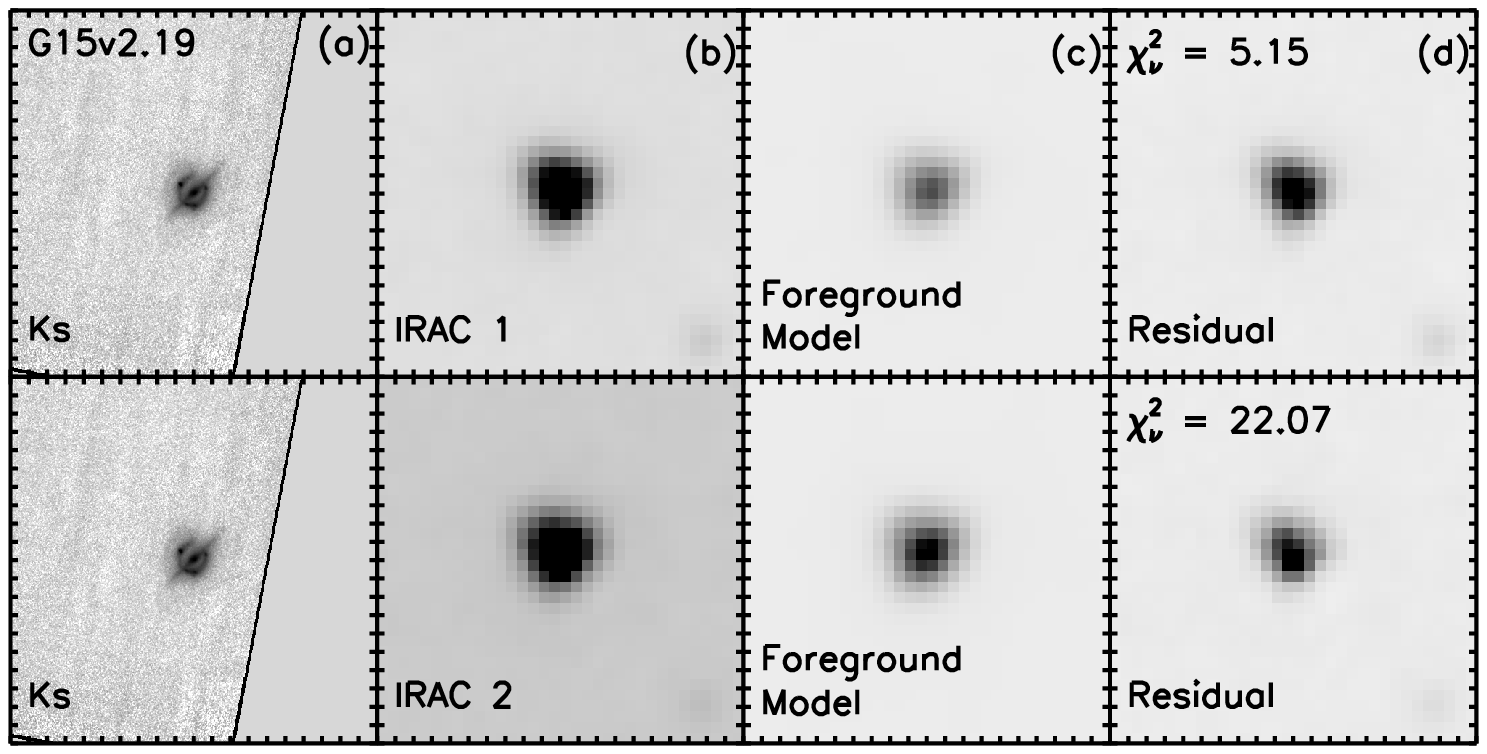} 
		\includegraphics[trim = 2cm 1.2cm 0.5cm 3.8cm,clip]{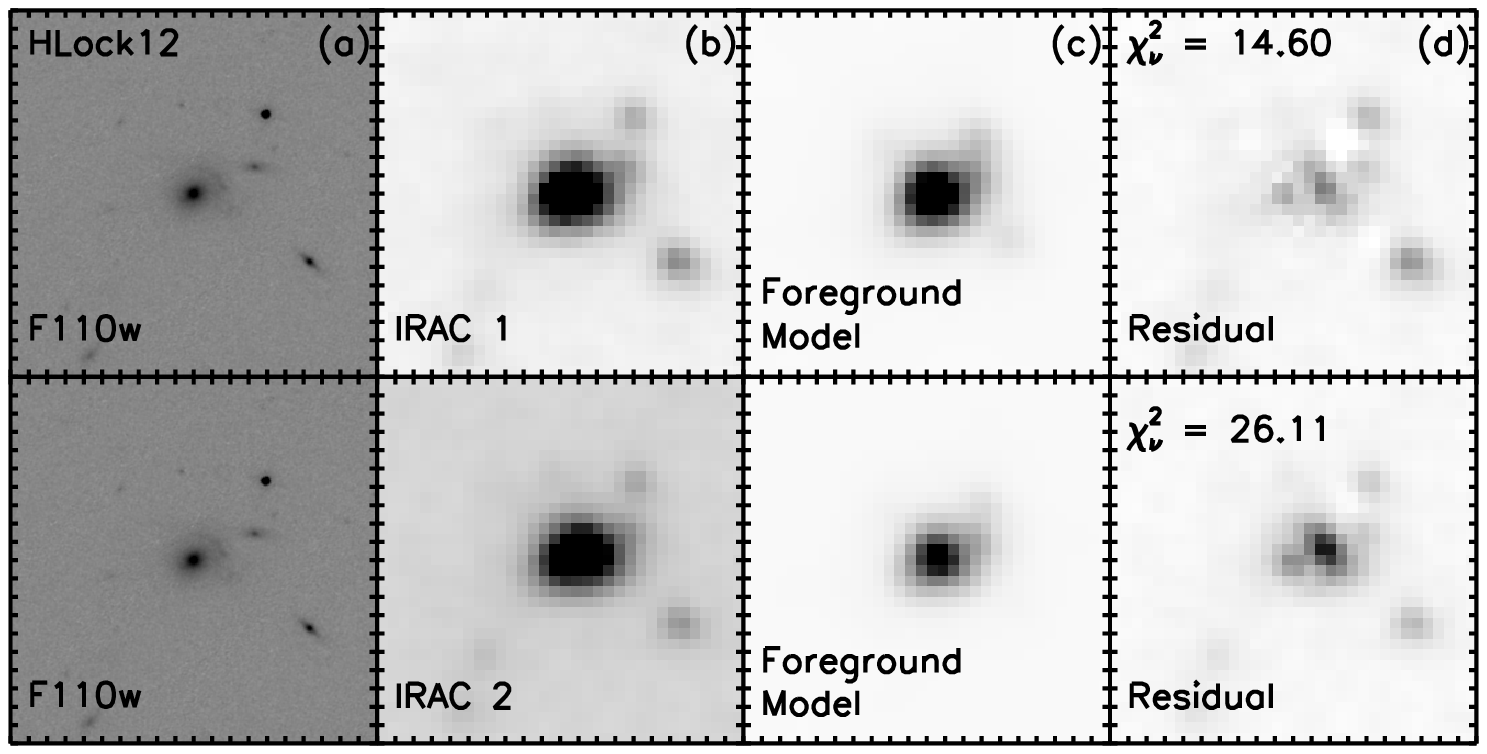}
	\end{center}
	\centerline{{\small \textbf{Figure~\ref{fig:deblend}} --- continued.}}
\end{figure}

%\begin{figure}
%	\begin{center}
%		\includegraphics[trim = 0 0 0 4cm]{./Figures/histogram}
%	\end{center}
%	\caption{Probability distribution functions (PDFs) of various {\sc Magphys} output parameters for each DSFG: (a) stellar mass and (b) star-formation rate. {\bf this has to be
%sampled with more points}}
%\end{figure}

\end{document}